\documentclass[10pt] {article}

\usepackage[usenames,dvipsnames,svgnames,x11names]{xcolor}
\usepackage{color}
\definecolor{darkblue}{rgb}{0.0,0.0,0.6}
\definecolor{maroon}{rgb}{0.5,0,0}
\definecolor{darkgreen}{rgb}{0,0.6,0}
\usepackage{bbding}
\usepackage{latexsym,amsthm,amsfonts,amsmath,mathrsfs}

\newcommand{\claim}[1]{\noindent\textcolor[rgb]{0.5,0.5,0.5}{$\blacktriangleright$}~\textit{{#1.}}~}

\newif\ifsectionbreaks
\sectionbreaksfalse

\usepackage[T1]{fontenc}
\usepackage[normalem]{ulem}

\usepackage[export]{adjustbox}

\usepackage{listings}

\lstdefinelanguage{XML}
{
basicstyle=\ttfamily\footnotesize,
  morestring=[b]",
  moredelim=[s][\bfseries\color{Maroon}]{<}{\ },
  moredelim=[s][\bfseries\color{Maroon}]{</}{>},
  moredelim=[l][\bfseries\color{Maroon}]{/>},
  moredelim=[l][\bfseries\color{Maroon}]{>},
  morecomment=[s]{<?}{?>},
  morecomment=[s]{<!--}{-->},
  commentstyle=\color{gray},
  stringstyle=\color{blue},
  identifierstyle=\color{red}
}

\usepackage{moreverb}
\usepackage[nounderscore]{syntax}

\usepackage{wrapfig}
\graphicspath{{./figures/}}
\DeclareGraphicsExtensions{.pdf}
\usepackage[export]{adjustbox}

\usepackage{amssymb}

\usepackage[
 font=footnotesize
 ]{subfig} 
 
\usepackage{algorithmicx}
\usepackage{algpseudocode}
\usepackage[ruled]{algorithm}
\usepackage{setspace}
\definecolor{light-gray}{gray}{0.75}
\algrenewcommand{\algorithmiccomment}[1]{\hskip3em{{\footnotesize \textcolor{light-gray}{$\blacktriangleright$}}} #1}

\usepackage{multirow} 
\usepackage{rotating} 
\usepackage{booktabs} 
\usepackage{colortbl} 
\usepackage{tablefootnote} 
\usepackage{array} 
\newcolumntype{L}[1]{>{\raggedright\let\newline\\\arraybackslash\hspace{0pt}}m{#1}}
\newcolumntype{C}[1]{>{\centering\let\newline\\\arraybackslash\hspace{0pt}}m{#1}}
\newcolumntype{R}[1]{>{\raggedleft\let\newline\\\arraybackslash\hspace{0pt}}m{#1}}

\usepackage[pdftex,colorlinks=true,urlcolor=blue,citecolor=blue]{hyperref}

\usepackage{enumitem}

\usepackage{blindtext}

\newcommand{\mobilenet}{MobileNet\xspace}
\newcommand{\resnet}{ResNet\xspace}
\newcommand{\yolo}{YOLO\xspace}
\newcommand{\lstm}{LSTM\xspace}
\newcommand{\bert}{BERT\xspace}

\begin{document}

\title{Pagoda: An Energy and Time Roofline Study for DNN Workloads on Edge Accelerators}

\author{Prashanthi S. K., Kunal Kumar Sahoo, Amartya Ranjan Saikia,\\ Pranav Gupta, Atharva Vinay Joshi, Priyanshu Pansari\\ and Yogesh Simmhan\\
Department of Computational and Data Sciences,\\
Indian Institute of Science, Bangalore 560012 India\\
Email: \{prashanthis, simmhan\} @iisc.ac.in}

\date{}

\maketitle

\begin{abstract}
Edge accelerators such as Nvidia Jetsons are becoming an integral part of the computing continuum, and are often used for DNN inferencing and training.
Nvidia Jetson edge devices have $2000$+ CUDA cores within a $70$W power envelope and offer $1000$s of power modes to customize CPU, GPU and memory frequencies. Their widely varying power--performance trade-offs can be exploited for energy and power-constrained deployments. While data-driven methods to predict the power and latency of DNN workloads for edge devices exist, there is a lack of principled study to understand \textit{why} edge accelerators and their power modes perform the way they do. We develop a \textit{time roofline} and a novel \textit{energy roofline} model for the Jetson Orin AGX for diverse power modes, and couple it with an \textit{analytical model} of the compute (FLOP) and memory access (bytes) for DNN inference workloads to analyze them from first principles. These reveal unique, sometimes counter-intuitive, insights into the power and performance behavior of DNN workloads on edge accelerators, e.g.,
the default power mode MAXN is not the most energy efficient and time efficiency implies energy efficiency for all power modes.
We also extend our analytical roofline models to DNN training.
Finally, we apply these methods to tune the power mode (and hence the roofline) of the edge device to optimize the latency and energy for DNN inference, with up to $15\%$ lower energy and minimal degradation in inference time.
\end{abstract}

\section{Introduction }
\label{sec:intro}

Edge accelerators are becoming a first-class computing platform 
as part of the computing continuum, complementing HPC and Cloud resources~\cite{continuum-jpdc-2022}.
They are being used to run Deep Neural Networks (DNNs) workloads, which are in widespread adoption for domains such as scientific computing~\cite{dnn-sci2}, healthcare~\cite{healthcare_DL} and smart cities~\cite{smartcity_DL}.
Besides low-latency data acquisition and inferencing close to instruments and field sensors~\cite{fmri-orin-2024,Yang_Lupascu_Meel_2021}, accelerated edge devices such as Nvidia Jetson are competitive for DNN training as well, to support continuous retraining over evolving data~\cite{ekya-stoica-nsdi} and privacy-oriented federated learning~\cite{hpdc-2020-tifl}. 
E.g., the latest AGX Orin Jetson device features an Ampere GPU with $2048$ CUDA cores, $12$-core Arm A78AE CPU at $2.2$GHz and $64$GB of RAM shared between GPU and CPU, with a performance of $275$ TOP/s (trillion ops per sec), making it comparable to a GPU workstation on performance, while consuming just $72$W of power with a compact size of $11$cm$\times$$11$cm$\times$$7$cm. Such edge hardware has sufficient capability to train and fine-tune a variety of models, including CNNs such as \resnet and \mobilenet, and transformer models such as \bert locally.

\subsubsection*{Challenges}
Edge devices are often deployed in field settings with power or energy constraints. E.g., they may be placed in forests to detect wildfires~\cite{Yang_Lupascu_Meel_2021}, where the high ambient temperature may restrict power load, or onboard a drone that may impose an energy constraint~\cite{ton-2024-drone}. Jetson edge devices provide knobs for users to tune the CPU, GPU, and memory frequency along with active CPU cores~\cite{ripeanu-2019-edge}, exposing over $18,000$ \textit{power modes} that can significantly impact power and performance~\cite{PowerTrain}. However, \textit{analyzing and explaining} the impact of these power modes on the performance and energy usage for DNN workloads is non-trivial~\cite{prashanthi2023sigmetrics,benchmarking_dnn_for_edge}. 
While data-driven and empirical methods can predict the behavior of DNN models for different power modes~\cite{PowerTrain,prashanthi2023sigmetrics}, they do not offer \textit{explainability} of the hardware characteristics or workload behavior for a power mode. This requires a methodical understanding
of the performance and energy characteristics.

Training and inferencing workloads execute multiple layers of the DNNs, with forward and backward passes that stress different resources of the hardware, e.g., CPU, GPU and memory. Batch sizes and precision also impact this behavior. Depending on the workload, \textit{distinct resources may be a bottleneck} that slows down the workload and impacts its power load. However, changing the power modes in a manner that over-provisions a resource that is not a bottleneck can cause the power to increase without performance benefits. E.g., changing the power mode to increase the CPU clock frequency when GPU is the bottleneck is not helpful.

Using the power modes to configure a \textit{balanced system} can ideally provide the best performance (least time) for a workload and with the least power load, leading to low energy usage. This concept is common in computer systems, e.g., Amdahl's Second Law for a balanced system~\cite{das:cloud:2017} and Bell et al.'s proposal for a balanced petascale computing system~\cite{gray-2006-computer}. 

\textit{Roofline models}~\cite{roofline_cacm} offer such an analytical tool to establish the \textit{balance point} that fully utilizes the compute capacity and memory bandwidth of the system to maximize the performance of a workload.
This can be particularly useful to configure the power modes of edge accelerators in a principled manner.

\subsubsection*{Gaps}
Roofline models~\cite{roofline_cacm} have been widely used in the HPC community to characterize and optimize scientific applications~\cite{roofline_HPC} and determine performance bottlenecks~\cite{carm_fgcs}. It has also been used on GPU servers to analyze the performance of DNN workloads~\cite{roofline_DL}. 
However, it is under-explored for deep learning on edge devices. PRoof~\cite{proof_icpp24} provides a low-overhead profiling technique based on analytical modeling for the time roofline, and considers less powerful and older Jetson edge devices and GPU servers. Their goal is to avoid extensive profiling to obtain the time roofline, whereas our focus is to develop both time and energy rooflines for contemporary edge accelerators to analyze their performance for deep learning. Others~\cite{matei_europar} characterize time rooflines on even older Jetsons, TK1 and TX1 (2015), for numerical and not DNN workloads.

Several works predict the power and performance of DNN workloads on servers~\cite{paleo_ICLR, NP_ACML, habitat_atc21}, but do not consider the impact of CPU, GPU and memory frequencies, which can be actively modified on Jetson edge devices. While there are some studies on the edge that consider these knobs~\cite{aslan:power}, including ours~\cite{PowerTrain, prashanthi2023sigmetrics}, they use ML-based prediction methods that fail to offer insights on \textit{why} a certain configuration is optimal for a given workload and why this differs across workloads.

\subsubsection*{Contributions} 
\textit{To the best of our knowledge, this work is the first to develop an energy roofline for edge accelerators, and study DNN training and inference workloads in the context of time and energy rooflines and diverse power modes for current edge devices such as the AGX Orin.}

While the roofline approach itself is well-established, our novel contribution lies in its use to analyze and quantify anticipated compute and power behavior of DNN workloads on edge accelerators. More importantly, we reveal \textit{counter-intuitive insights}, e.g., the fastest (and default) power mode is not the most energy-efficient; GPU frequency impacts not just FLOP/s but also memory bandwidth~\footnote{In this article, we use FLOP/s when referring to performance of the hardware given by \textit{Floating Point Operations per second}, and FLOP when referring to the total computing needs of a workload given by \textit{Floating Point Operations}.}; a DNN workload can simultaneously be compute-bound in energy and memory-bound in time; and optimizing for time automatically optimizes for energy. 
These takeaways are useful for practitioners to correlate model performance to device specifications, choose or configure devices and DNN workloads to obtain the best performance, and help both ML researchers and hardware designers design better models and improve architectural bottlenecks.

We make the following specific contributions: 
\begin{enumerate}[leftmargin=*]
\item We model \textit{time and energy rooflines} for the Nvidia Jetson AGX Orin using empirical evaluation of various power modes (\S~\ref{sec:methodology:roofline}). We analyze them to understand the effect of various GPU, CPU and memory frequencies on the floating point and memory bandwidth performances (\S~\ref{subsec:results_time_roof}, \S~\ref{subsec:results_energy_roof}). We also identify the time and energy \textit{balance points} in the roofline and examine the relationship between time and energy efficiency. 

\item We develop an \textit{analytical model} to estimate the FLOP and memory accesses required for $5$ diverse \textit{DNN inference workloads}: \resnet, \mobilenet, \yolo, \lstm and \bert (\S\ref{sec:methodology:analytical}). We analyze their performance in the context of the roofline for various power modes, and also the the effect of two common workload knobs, arithmetic precision and batch size, on the \textit{arithmetic intensity} and performance (\S\ref{subsec:ncu_val_w_analytical_model}). 

\item We briefly extend our analytical model to $5$ DNN \textit{training workloads} and offer early analysis on their performance in the context of the time and energy roofline (\S\ref{sec:training}). 

\item As an application of the inference roofline, we examine whether \textit{shifting the time roofline} according to the workload's arithmetic intensity can help reduce power without degrading performance, resulting in energy benefits (\S\ref{sec:casestudy}). 
\end{enumerate}

Apart from this, we present a background of edge accelerators and DNN workloads (\S\ref{sec:motivation}), related work on edge characterization and roofline studies (\S\ref{sec:related}) and details of our experimental setup (\S\ref{sec:setup}).

While this study is limited to the AGX Orin, the methodology is applicable to other edge platforms, and complementary studies~\cite{PowerTrain} have demonstrated generalization across diverse edge hardware. Similarly, it can also extend to Large Language Models (LLMs) and other emerging architectures~\cite{mayank:ipdpsw:2025}.

\section{Background and Motivation}
\label{sec:motivation}

\begin{figure}[t]
\centering
\captionof{table}{Nvidia Jetson Orin AGX 32GB Dev Kit Specifications}
\vspace{-0.1in}
\footnotesize
\setlength{\tabcolsep}{6pt}
\renewcommand{\arraystretch}{1.2}
\label{tbl:hwspecs}
\begin{tabular}{p{3cm}|p{2cm}||p{3cm}|p{0.8cm}}
\hline
\textbf{Hardware Specs.} & \textbf{Value} & \textbf{Power Modes Specs.} & \textbf{Value} \\ \hline\hline
CPU Architecture & ARM A78AE & \# CPU freqs. & 29\\ \noalign{\global\arrayrulewidth=0.4pt}\arrayrulecolor{lightgray}\hline
CPU Cores & 12 & CPU cores & 1--12\\ \noalign{\global\arrayrulewidth=0.4pt}\arrayrulecolor{lightgray}\hline
  &  & Max CPU Freq.   & 2.2GHz\\ \noalign{\global\arrayrulewidth=0.4pt}\arrayrulecolor{black}\hline
GPU Architecture & Ampere & \# GPU freqs. & 13\\ \noalign{\global\arrayrulewidth=0.4pt}\arrayrulecolor{lightgray}\hline
CUDA, Tensor Cores & 2048, 64 & Max GPU Freq. & 1.3GHz\\ \noalign{\global\arrayrulewidth=0.4pt}\arrayrulecolor{black}\hline
RAM (GB) & 32 & \# Mem freqs. & 4\\ \noalign{\global\arrayrulewidth=0.4pt}\arrayrulecolor{lightgray}\hline
RAM Type & LPDDR5 & Max Mem Freq. & 3.2GHz\\ \noalign{\global\arrayrulewidth=0.4pt}\arrayrulecolor{black}\hline
Accelerators & DLA & DLA Max Freq. & 1.4GHz\\ \noalign{\global\arrayrulewidth=0.4pt}\arrayrulecolor{lightgray}\hline
 & PVA & PVA Max Freq. & 0.7GHz\\
 \noalign{\global\arrayrulewidth=0.4pt}\arrayrulecolor{black}\hline
Form Factor (mm) & 110$\times$110$\times$72 & Max Power (W) & 72\\ \noalign{\global\arrayrulewidth=0.4pt}\arrayrulecolor{lightgray}\hline
Price (USD) & \$1999 & Total \# Power Modes & 18,086\\ \noalign{\global\arrayrulewidth=0.4pt}\arrayrulecolor{black}\hline
\end{tabular}
\vspace{-0.2in}
\end{figure}

\subsection{Jetson Edge Accelerators and Power Modes}
The Jetson AGX Orin developer kit is the latest and most powerful edge device from Nvidia~\cite{OrinAI_benchmarks}, popular for edge AI workloads. Besides $4$ standard power modes with a target power budget, these devices allow users to define custom power modes through fine-grained control over the CPU cores, and CPU, GPU, and memory frequencies. As shown in Table~\ref{tbl:hwspecs}, 
we have $\approx 18,000$ custom power modes with vastly different power--performance trade-offs. This is particularly relevant in field deployments of edge devices across the computing continuum that may be constrained in power or energy, and need to optimize performance within those constraints. E.g., \bert inference with a batch size of $32$ using the MAXN default power mode, which offers the highest performance, consumes $62.2$W of power and takes $1919$ms per batch while running it on a lower power mode with $4$ CPU cores, $0.4$GHz CPU, $0.1$GHz GPU, and $0.6$GHz memory increases this inference time to $16,677$ms and decreases the power to $15.6$W -- a difference of $\approx 8.7\times$ in time and almost $3.9\times$ in power. Such diversity is confirmed by prior empirical and data-driven studies~\cite{prashanthi2023sigmetrics, ripeanu-2019-edge}. However, a principled analytical modeling of these devices using a time and energy roofline study is lacking 
for DNN workloads.

\subsection{Roofline Model and Arithmetic Intensity}

\begin{figure}[t]
\centering
\subfloat[Time Roofline]{
    \includegraphics[width=0.50\columnwidth,trim={0 0.5cm 0 0},clip]{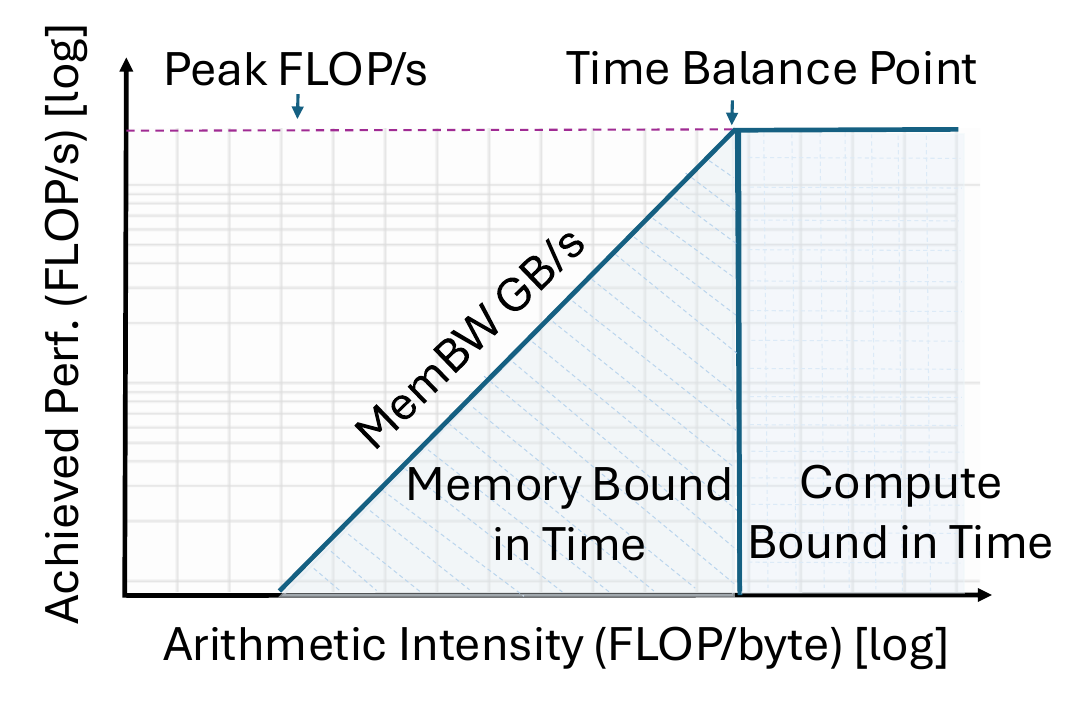}
    \label{fig:roofline_time}
  }
\subfloat[Energy Roofline]{
    \includegraphics[width=0.46\columnwidth]{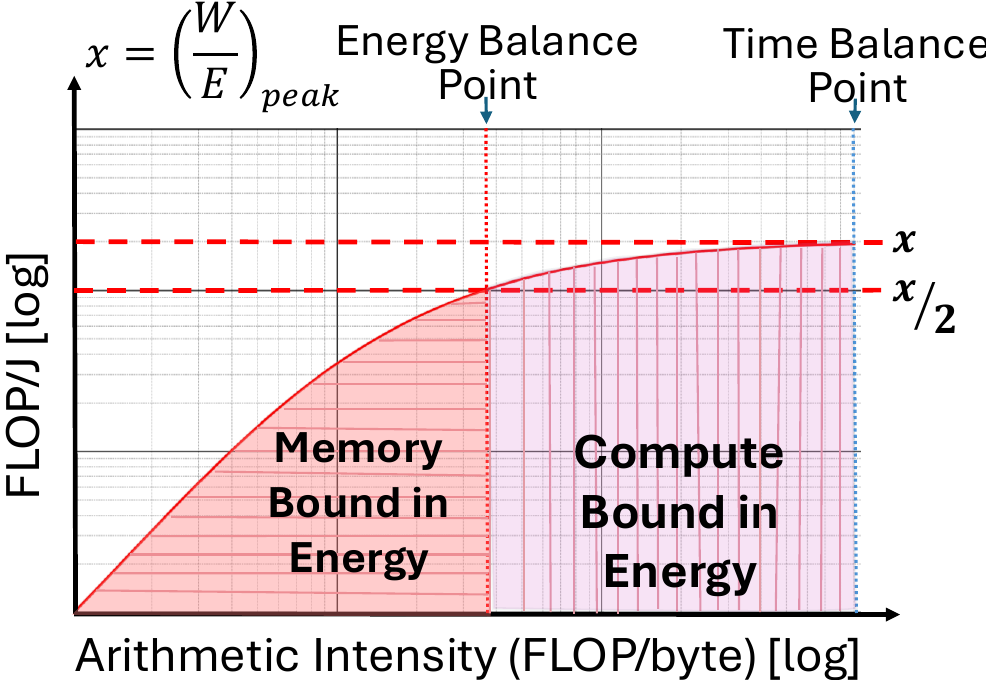}
    \label{fig:roofline_power}
  }

\caption{Roofline Model Examples} 
\label{fig:roofline}
\end{figure}

The time roofline model~\cite{roofline_cacm} was proposed as a visual performance model for computing hardware, which allows programmers and system designers to identify bottlenecks and performance optimizations. While it was originally proposed for CPUs, it has since been widely used and extended to GPUs~\cite{gpu_roofline}. Fig.~\ref{fig:roofline} shows an exemplar  \textit{time roofline}, and reports the \textit{performance achieved} by a workload (FLOP/s) having a specific \textit{Arithmetic Intensity (AI)}~\footnote{We use \textit{AI} to refer to Arithmetic Intensity in this article and not Artificial Intelligence.} jointly as a function of the \textit{memory and compute performance} of the hardware.
The arithmetic intensity is the ratio of the total compute (FLOP) to total memory operations (MOP) for an application and is a measure of how much compute is performed for unit data. 
Workloads such as dense matrix multiplications typically have high AI and exhibit efficient data reuse, while sparse matrix-vector multiplications have irregular memory accesses and low AI~\cite{roofline_cacm}.

The performance of a workload is limited either by peak memory bandwidth or peak compute power of the hardware. Workloads with low AI are memory-bound since the peak memory bandwidth of the device is insufficient to keep the processor busy. In this \textit{memory-bound region}, the achieved application performance (FLOP/s) increases as AI for the workload increases. As AI grows, the compute gets more utilized for each byte fetched from memory and the overall application performance increases. Beyond a limit, known as the \textit{balance point} (or ridge point), the AI is high enough that it saturates the compute capacity and we are in the \textit{compute-bound region}. The processor is unable to keep up with the speed at which memory delivers data. When the workload's AI matches the balance point of the system, it is making full use of both memory bandwidth and processing capacity, and is bottlenecked by neither.

There are several roofline variants, such as the hierarchical roofline~\cite{yang2021hierarchical}, cache-aware roofline, etc.~\cite{marques2020application}. 
A relevant one for us is the \textit{energy roofline}~\cite{roofline_energy_ipdps} (Fig.~\ref{fig:roofline_power}) that provides the \textit{peak energy efficiency achieved} by a workload as a function of its AI. The \textit{energy balance point} is one where the workload's AI causes equal amounts of energy to be spent on computations and memory transfers.

 \vspace{-0.1in}
\subsection{DNN Inference and Training on Edge}
Applications choose a DNN model by a combination of its accuracy, power and performance~\cite{dnn_benchmark}.
DNN inference is more lightweight and involves just the forward pass, while training also involves a backward pass followed by a weight update in order to iteratively refine the model. On Jetson edge devices, both training and inference use the GPU for computation, and preprocessing and kernel launches are done by the CPU~\cite{prashanthi2023sigmetrics}. The same memory is shared between CPU and GPU, and memory is accessed for reading/writing the weights, inputs and intermediate outputs. Thus, the power load and performance of DNN workloads depend on CPU, GPU, and memory resources. The roofline model offers an intuitive and generalizable method to model the device performance and its behavior for DNN workloads with varying AIs. Further, since Jetson devices offer fine-grained power modes, each power mode results in a different roofline behavior. Therefore, it is crucial to analyze DNN power and performance within this framework.

\vspace{-0.1in}
\section{Related Work}
\label{sec:related}

\subsubsection*{Roofline studies for DNN workloads}
PRoof~\cite{proof_icpp24} proposes a profiling-based roofline approach along with an analytical model of FLOP and memory of DNN inference workloads. Their main focus is to eliminate the need for time-consuming hardware profiling required to build the roofline and replace this with an analytical model. They validate this on GPU servers and older/slower Jetsons. We leverage their time roofline method but with the goal of analyzing DNN inference performance across various batch sizes, power modes and precisions for the latest Orin AGX device. Further, we also develop energy rooflines for the edge device and extend the analysis to DNN training workloads. Others~\cite{matei_europar} have characterized the roofline of older Jetsons, TK1 and TX1, to evaluate the impact of frequency scaling on the roofline. Their study is limited to microbenchmarks and does not consider DNN workloads, which are complex. Roofline models have also been used to analyze LLM inferencing~\cite{llminf_roofline} and specific optimizations such as quantization and speculative decoding. In our work, we consider a broader variety of DNN inference workloads, including CNNs, RNNs and transformer models.

\subsubsection*{Roofline model and applications in HPC}
The roofline model~\cite{roofline_cacm} was proposed as a method to draw insights on possible performance optimizations for applications. Others~\cite{gpu_roofline} have extended it to the GPUs. The roofline has also been widely used in the HPC community to characterize and optimize scientific applications~\cite{roofline_HPC} and determine bottlenecks~\cite{carm_fgcs}. A prior study~\cite{roofline_energy_ipdps} has also proposed an energy roofline model to help understand time--energy tradeoffs. While these works propose different variants of the roofline model, we leverage the traditional time and energy roofline~\cite{roofline_cacm, roofline_energy_ipdps} to analyze the performance and energy of DNN workloads on edge accelerators.

\subsubsection*{Power and performance prediction for DNN workloads on edge devices}
Power and runtime latency predictions for GPU servers use polynomial regression~\cite{NP_ACML} 
to predict the runtime and power of inference workloads, and
Multi-Layer Perceptrons and wave scaling~\cite{habitat_atc21} for DNN training. These do not consider power modes, which are a unique feature of edge devices. There has been some work on edge devices that use ML-based prediction techniques such as linear regression~\cite{prashanthi2023sigmetrics} and transfer learning with neural networks~\cite{PowerTrain} for power and performance prediction of DNN training across power modes. However, these methods are data-driven and do not offer a fundamental understanding into why a certain power mode may be optimal for a given workload. In contrast, we develop and use methods based on first principles to combine an analytical modeling of the DNN workload with the roofline of the edge device across power modes.

\subsubsection*{Analytical modeling of compute and memory}
Paleo~\cite{paleo_ICLR} predicts the DNN training time on GPU servers by modeling the layerwise compute and memory times, and empirically estimating a scaling constant for the peak achievable performance. Others~\cite{cnninf_pmlr} model the FLOP and memory access for CNN inference workloads but use this to enhance the CNN architecture to have higher AI. We also theoretically model the compute and memory of DNN inference models, but with the goal of analyzing the positioning of their AI in the time and energy roofline models and tuning the power modes to shift the rooflines rather than the workload's AI. 

\vspace{-0.1in}
\section{Experiment Setup}
\label{sec:setup}

\subsection{Hardware}
We use the Nvidia Jetson Orin AGX developer kit's $32$GB variant for the experiments and analysis in this article (Table~\ref{tbl:hwspecs}). It has a 12 core ARM A78AE CPU with a peak of $2.2$GHz, an Ampere GPU with $2048$ CUDA cores and $64$ tensor cores at a maximum of $1.3$GHz, and $32$GB of LPDDR5 RAM shared between the CPU and GPU, clocked at a maximum frequency of $3.2$GHz. The tensor cores are intended to accelerate matrix multiplications and support Tensor Float 32 (TF32)~\cite{TF32}, a proprietary floating-point format introduced by Nvidia, apart from FP16 and INT8. Our workloads utilize both the CUDA cores and the Tensor cores. The developer kit also features two additional accelerators DLA and PVA. The PVA and DLA  accelerators are not used in our experiments as the PVA does not support DNN workloads and the DLA supports only a restricted set of DNN workloads but with GPU fallback. 

The default power mode is MAXN, which sets all resources at their peak performance. It also has 3 other preset power modes with a target power load. The user can also custom-configure the CPU/GPU/memory frequencies and the CPU cores from a set of supported values, giving $18,086$ possible power modes. Table~\ref{tbl:hwspecs} lists the minimum, maximum, and number of possible values for each resource in a power mode. We use the default MAXN power mode for all experiments unless stated otherwise, as it represents the peak performance configuration. However, we collect roofline data for $96$ diverse power modes to understand the effect of varying GPU, CPU, and memory frequencies on performance and energy efficiency. In addition, we explicitly explore the use of different power modes in our analysis on shifting the roofline to optimize DNN inference. Our study is not limited to MAXN, but rather uses it as a baseline to demonstrate the insights gained from analyzing a wide range of power modes.

\vspace{-0.1in}
\subsection{Software and System Configuration}
The Orin AGX is flashed with JetPack v$5.0.1$, the latest available JetPack at the start of this study, and this version of the Jetson operating system is officially configured with CUDA v$11.4$ and Ubuntu $20.04$ LTS. We use PyTorch as the framework for all our inference workloads, with torch v$1.12.0$ and torchvision v$0.13.0$. All datasets are stored on the SSD, and we use the PyTorch Dataloader with $4$ workers to fetch data. We use the default MAXN power mode for all experiments, unless stated otherwise. We use FP32 as the default precision for all DNN workloads, but since Ampere, these are automatically processed using TF32 on tensor cores wherever possible~\cite{TF32}. We use a default batch size of $bs=1$ for inference workloads, but vary it in specific experiments. For training, we do not vary batch size as it is a hyperparameter and fix the batch size at $bs=16$. 

Dynamic Voltage and Frequency Scaling (DVFS) is disabled to ensure that frequency values stay constant for a power mode. All 8 GPU Texture Processing Clusters (TPCs) are left on with only the GPU frequency changing.
We do not alter the frequencies of the DLA and PVA from their default values since they are not used. 
The fan is set to maximum speed to avoid overheating and thermal throttling. The onboard power sensor is used for power measurements, which captures the power of the module and not the carrier board and peripherals.

\begin{table}[t]
\centering
\footnotesize
\setlength{\tabcolsep}{1pt}
\renewcommand{\arraystretch}{1.2}
\caption{DNN Models and Datasets used in Experiments.} 
\label{tbl:modeldataset}
\begin{tabular}{p{1.9cm}|p{2cm}||p{3cm}|p{1cm}|p{1cm}||p{3cm}}

\hline
\bf{Task} & \bf{Mode} & \bf{Model} & \bf{Layers} & \bf{Params} & \bf{Dataset} \\ \hline\hline
Image classif. & Infer. & \textbf{\resnet-50~\cite{resnet}}& 50  & $25.6M$ & \textbf{ImageNet~\cite{imagenet}} \\ \hline
Image classif. & Train. & \textbf{\resnet-18~\cite{resnet}}& 18  & $11.7M$ & \textbf{ImageNet~\cite{imagenet}} \\ \hline
Image classif. & Infer., Train. & \textbf{{\mobilenet}v3~\cite{mobilenet}}& 20  & $5.5M$ & \textbf{GLD23k~\cite{tensorflow_gld23k}} \\ \hline
Object detect. & Infer., Train. &\textbf{\yolo-v8n~\cite{yolo}} & 53 &  $3.2M$  &  \textbf{COCO mini.~\cite{coco_minitrain}} \\ \hline
Next word & Infer., Train. & \textbf{LSTM~\cite{LSTM}}& 2  & $8.6M$  & \textbf{Wikitext~\cite{wikitext}} \\ \hline
Question ans. & Infer. & \textbf{BERT Large~\cite{bert_large}}& 24 & $340M$ & \textbf{SQuAD V2.0~\cite{squad}}  \\ \hline
Question ans. & Train. & \textbf{BERT Base~\cite{bert_large}}& 12 & $110M$ & \textbf{SQuAD V2.0~\cite{squad}}  \\ \hline
\end{tabular}
\end{table}

\vspace{-0.12in}
\subsection{Models and Datasets}
For inference, we choose five representative DNN workloads -- MobileNet v3~\cite{mobilenet}, ResNet-50~\cite{resnet}, YOLO-v8n~\cite{yolo}, LSTM~\cite{LSTM} and BERT Large~\cite{bert_large} -- shown in Table~\ref{tbl:modeldataset} along with their datasets. These encompass a broad spectrum of tasks common in real-world edge applications~\cite{AbelmoniemREFL, jallepalli_federated_yolo}. They also exhibit significant variability in their model architectures, spanning CNNs, RNNs and transformers, and with a wide range of sizes, from a compact 3.2M parameters to more complex 340M parameters. The compute demands of these workloads also vary considerably, ranging from 130M -- 79B estimated FLOP.
For training, we use the same model wherever possible (\mobilenet, \yolo, \lstm), and smaller variants of models when memory footprint is a limitation (\bert Base and \resnet-18).

\vspace{-0.1in}
\subsection{Profiling Setup}
For batch inference/training time profiling, we instrument the PyTorch code with \texttt{torch.cuda.event} with the \texttt{synchronize} flag to accurately measure execution time on the GPU. We run for $\approx 40$ batches for each power mode, and omit the initial batch values to ignore startup overheads. For power measurements, we use the \texttt{jtop} utility, which is a wrapper around Nvidia tegrastats. The power readings are sampled at 1s intervals. The power values take $2$--$3$s to stabilize after the workload starts and we account for that. The energy is calculated as a sum of the sampled power, weighted by time intervals between consecutive samples.

\begin{table}[t]
\centering
\footnotesize
\setlength{\tabcolsep}{1.5pt}
\renewcommand{\arraystretch}{1.1}
\def\thickhline{\noalign{\hrule height1pt}}
\caption{Key Notations Used}
\label{tbl:notation}
\begin{tabular}{c|l}
\hline
    \textbf{Symbol} & \textbf{Description} \\
  \thickhline
    $W$ & Number of Arithmetic Operations (FLOP)\\ \hline
    $Q$ & Number of Bytes of Memory Operations (MOP) \\ \hline
    $I = W/Q$ & Arithmetic Intensity (AI)\\ \hline
    $\text{FLOP/s}_\text{(peak)}$ & Peak Compute Performance for a power mode\\ \hline 
    $\text{BW}_{\text{(peak)}}$ & Peak Bandwidth for a power mode\\ \hline
    $\tau_\text{flop} = 1/\text{FLOP/s}_\text{(peak)}$ & Time taken per FLOP for a power mode\\ \hline
    $\tau_\text{mop} = 1/\text{BW}_{\text{(peak)}}$ & Time taken per MOP for a power mode\\ \hline
    $\epsilon_\text{flop}$ & Energy consumed per FLOP for a power mode\\ \hline
    $\epsilon_\text{mop}$ & Energy consumed per MOP for a power mode\\ \hline
    $\beta_\tau$ & AI at Time balance point\\ \hline
    $\beta_\epsilon$ & AI at Energy balance point\\ \hline
    $E$ & Energy (J) consumed by the workload on the device\\ \hline
    
\end{tabular}
\vspace{-0.2in}
\end{table}

\vspace{-0.1in}
\section{Proposed Roofline Models}
\label{sec:methodology:roofline}

We provide a brief overview of the time and energy roofline models and explain how we empirically obtain them for the Orin AGX. We also present derivations for the energy roofline with and without static (base load) power, expanding upon the seminal work for servers~\cite{roofline_energy_ipdps}. A summary of the notation used is in Table~\ref{tbl:notation}, and kept consistent with
the prior work~\cite{roofline_energy_ipdps}.

\subsection{Time Roofline}
The time roofline gives an upper bound of the achievable workload performance as its arithmetic intensity varies. $W$ is the number of arithmetic operations performed by the workload (FLOP) and $Q$ is the number of Bytes of Memory Operations (MOP).
The \textit{Arithmetic Intensity (AI)} of a workload is $I = \frac{W}{Q}$. The time per FLOP and MOP are given by
$\tau_\text{flop} = \frac{1}{\text{FLOP/s}_\text{(peak)}}$ and $\tau_\text{mop}= \frac{1}{\text{BW}_\text{(peak)}}$. 
Here, \(\text{FLOP/s}_{\text{(peak)}}\) is the \textit{peak compute performance} of the device for the specific power mode and $\text{BW}_\text{(peak)}$ the \textit{peak memory bandwidth} from DRAM to the GPU cores. 

So, \( T_\text{flop} = W \cdot \tau_\text{flop} \) and \(T_\text{mop} = Q \cdot \tau_\text{mop}\) give the total time taken for compute execution and memory transfer operations. The roofline model assumes that compute and memory operations occur in parallel. So the \textit{total runtime of a workload} having an AI of $I$ is:
\begin{align*}
    T &= \max\left( T_\text{flop}, T_\text{mop} \right) 
      = \max\left( W \cdot \tau_\text{flop}, Q \cdot \tau_\text{mop} \right) \\
      &= W \cdot \tau_\text{flop} \cdot \max\left( 1, \frac{\beta_\tau}{I} \right) 
      \label{eqn:time_roofline}
\end{align*}
\vspace{1pt}

The \textit{time roofline} plots the relation between $\frac{W}{T}$ and $I$.
The \textit{time balance point} (or ridge point) $\beta_\tau$ is when the time taken for compute and memory operations is identical, and the AI of the workload is in balance with the system. 
When the AI of the workload is less than the balance point, it is bottlenecked by memory bandwidth and is said to be memory bound. Likewise, when the workload's arithmetic intensity is greater than the balance point, it is compute bound.

\begin{equation}
T = \begin{cases}
    T_\text{flop} = \frac{W}{\text{FLOP/s}_\text{(peak)}}, & \text{compute-bound} \\
    T_\text{mop} = \frac{Q}{\text{BW}_\text{(peak)}}, & \text{memory-bound}
\end{cases}
\label{eq:time_split}
\end{equation}
\vspace{2pt}

\subsection{Energy Roofline}
The energy roofline gives an upper bound of the energy efficiency with varying AI. The total energy consumed is the sum of energy consumed due to static power, i.e., base load power even when the system is idle, and energy consumed for compute and memory operations. Since energy is additive (unlike compute and memory times that can be overlapped), the time roofline shows a sharp transition while the energy roofline is a smoother arch~\cite{roofline_energy_ipdps}.
If $\epsilon_{\text{flop}}$ and $\epsilon_{\text{mop}}$ are the energy per unit FLOP and MOP, and $\pi_0$ is the static power, the energy consumed by the workload $E$ is:

\begin{equation}
    E\left(W, Q, T\right) = \epsilon_{\text{flop}} \cdot W + \epsilon_{\text{mop}} \cdot Q + \pi_0 T  \label{eq:energy_all}
\end{equation}

Assuming only compute operations are useful work and memory operations are overheads, we define \textit{energy efficiency} as:
\begin{equation}
    \frac{W}{E} = \frac{W}{\epsilon_{\text{flop}}\cdot W + \epsilon_{\text{mop}}\cdot Q + \pi_0 T}  \label{eq:energy_efficiency}
\end{equation}
The \textit{energy roofline} plots the relation between $\frac{W}{E}$ and $I$.

\subsubsection{Peak Energy Efficiency, without static power} 
We first assume that the device consumes no static power $(\pi_0 = 0)$. From Eqn.~\ref{eq:energy_efficiency}:
\begin{align*}
    \frac{W}{E} &= \frac{W}{\epsilon_{\text{flop}} \cdot W + \epsilon_{\text{mop}} \cdot Q}
                = \frac{1}{\epsilon_{\text{flop}} + \epsilon_{\text{mop}} \cdot \frac{Q}{W}} 
                = \frac{1}{\epsilon_{\text{flop}} + \epsilon_{\text{mop}} \cdot \frac{1}{I}}
\end{align*}
\vspace{1pt}
As $I$ increases, the energy decreases and energy efficiency increases.
Since $I = \frac{W}{Q}$, we can either increase the FLOP, $W$, or decrease the MOP, $Q$, to increase AI, and achieve peak energy efficiency. When the workload's FLOP increases to infinity, the AI of the workload also tends to infinity. The peak energy efficiency is:
        \begin{align}
            \lim_{W \to \infty} \frac{W}{E} = \lim_{I \to \infty} \frac{1}{\epsilon_{\text{flop}} + \epsilon_{\text{mop}}\frac{1}{I}}
            = \frac{1}{\epsilon_{\text{flop}}} 
            \label{eq:peak_energy_efficiency}
        \end{align}
      \vspace{1em}

Similarly, when its MOP drops to zero, $I\to\infty$ and the peak energy efficiency is the same as the Eqn. \ref{eq:peak_energy_efficiency}. Hence, in both cases, in the absence of static power load, the energy efficiency saturates to the reciprocal of the energy needed to perform one compute operation.

The \textit{energy balance point} $\beta_{\epsilon}$ is the AI at which equal energy is spent on FLOP and MOP. So, the energy efficiency at this point is half of the peak energy efficiency~\cite{roofline_energy_ipdps}.

Formally:
\vspace{3pt}
\begin{equation}
    \beta_{\epsilon} = \frac{\epsilon_{\text{mop}} +\frac{\pi_0} {\text{BW}_{\text{(peak)}}} }{\epsilon_{\text{flop}} + 2\frac{\pi_0} {\text{FLOP/s}_{\text{(peak)}}}} = \left. \frac{\epsilon_{\text{mop}}}{\epsilon_{\text{flop}}} \right|_{\pi_0=0}
    \label{eq:energy_balance_non_zero_pi_not} 
\end{equation}
\vspace{4pt}

\subsubsection{Peak Energy Efficiency, with static power} 
Here, we include the static power consumed ($\pi_0>0$) and again vary AI either by increasing the FLOP or decreasing the MOP for the workload in Eqn.~\ref{eq:energy_efficiency}. 

As the number of compute operations $W$ increases, we may get into either the memory-bound or the compute-bound region of the time roofline.

In the \textit{memory-bound region}, the runtime is slowed by the time taken for memory operations (\(T = T_{\text{mop}} \)), which is a constant. We can rewrite Eqn.~\ref{eq:energy_efficiency} as:
        \begin{align*}
            \frac{W}{E} &= \frac{W}{W \cdot \left( \epsilon_{\text{flop}} + \epsilon_{\text{mop}} \cdot \frac{1}{I} + \pi_0 \cdot T_{\text{mop}} \frac{1}{W} \right)} \\
                        &= \frac{1}{\left( \epsilon_{\text{flop}} + \epsilon_{\text{mop}} \cdot \frac{1}{I} + \pi_0 \cdot T_{\text{mop}} \cdot \frac{1}{W} \right)}
        \end{align*}
        As $W$ increases, the denominator decreases, and the overall energy efficiency increases.
       
In the \textit{compute-bound region}, the runtime is dominated by the compute time (\(T = T_{\text{flop}} \))

         \noindent As before, when we increase $W$ to infinity to increase AI, the energy efficiency Eqn.~\ref{eq:energy_efficiency} becomes:
        
        \begin{align*}
            \lim_{W \to  \infty} \frac{W}{E} &= \lim_{W \to \infty} \frac{W}{\epsilon_{\text{flop}} W + \epsilon_{\text{mop}} Q + \pi_0 T_{\text{flop}}} = \frac{1}{\epsilon_{\text{flop}} + \frac{\pi_0}{\text{FLOP/s}_{\text{(peak)}}}}
        \end{align*}
   
    \noindent Similarly, when we decrease $Q$ to increase AI
    we achieve the same result as earlier. 
    Hence, the presence of static power lowers the peak and the overall energy efficiency. Also, as seen from Eqn. \ref{eq:energy_balance_non_zero_pi_not} the energy balance point, $\beta_{\epsilon}$, is lower for a device with static power compared to a device with no static power.

\subsubsection{Energy Roofline in terms of Time Roofline}
We observe that the energy roofline transitions at the time balance point. We analyze this by examining the energy equations before and after the time balance point, which has not been done in existing literature. 
    On dividing Eqn.~\ref{eq:energy_all} by W on both sides, we get :

    \begin{equation}
        \frac{E}{W} = {\epsilon_{\text{flop}}} + \frac{\epsilon_{\text{mop}} \cdot Q}{W} + 
        \frac{\pi_0  \cdot T}{W}  
        \label{eq:inverse_energy_efficiency}
    \end{equation}
       Substituting $T$ for the memory-bound region in Eqn.~\ref{eq:inverse_energy_efficiency}:
    \begin{equation}
        \frac{1}{W/E}
        = \epsilon_{\text{flop}} + \frac{\epsilon_{\text{mop}}}{W/Q} +
        \frac{\pi_0 / \text{BW}_{\text{(peak)}}}{W/Q} 
        \label{eq:Final EquationA}
    \end{equation}
   
    \noindent Similarly, for compute-bound, we get:

    \begin{equation}
        \frac{1}{W/E} = \epsilon_{\text{flop}} + \frac{\epsilon_{\text{mop}}}{W/Q} + 
        \frac{\pi_0 }{\text{FLOP/s}_{\text{(peak)}}}
        \label{eq:Final EquationB}
    \end{equation}

Both Eqns.~\ref{eq:Final EquationA} and~\ref{eq:Final EquationB} are of the form
        $\frac{1}{y} = A + \frac{B}{x}$
which represents a hyperbola, as visualized in Fig.~\ref{fig:roofline_power}. 

\subsubsection{Empirically Determining Roofline Coefficients}
\label{sec:emproofline}
We extend PRoof's microbenchmarks~\cite{proof_icpp24} to empirically obtain the coefficients for our time and energy rooflines equations above. For instance, for the MAXN power mode, the energy per flop is $\epsilon_\text{flop}=3.86$~pJ/FLOP, energy per memory operation is $\epsilon_\text{mop}141.38$~pJ/byte, and static power $\pi_0$ is $17.9$W. Our roofline workload is designed to be representative of DNN workloads since our goal is not just to empirically measure the peak performance and bandwidth but also validate the roofline's use as a proxy for DNN performance, and shape it accordingly. The \textit{compute-bound workload} consists of matrix multiplications, and the \textit{memory-bound workload} consists of ReLU activation and transpose operations, all of which are common operations in DNNs. For the memory workload, we do not include concat operations as they cause bandwidth measurements higher than the theoretical peak DRAM bandwidth due to heavy cache accesses. We find the largest matrix size that fits in memory and runs on the Jetson Orin, and vary the matrix sizes in powers of $2$ until this limit. For a given power mode, we measure the peak compute and memory performance across all these matrix sizes. We also record the time taken and the power consumed.
These are used to estimate $\text{FLOP/s}_{\text{(peak)}}$ and $\text{BW}_\text{(peak)}$ as a median over $10$ runs. 

By default, we use FP32 precision for our roofline workloads. However, as mentioned earlier, Jetsons automatically run these on tensor cores using TF32 wherever possible, which is the case for these DNN workloads. Later, we compare our roofline benchmarks with Nvidia's benchmarks~\cite{CUDA_benchmarks} for compute and memory, and show that the results match. 

We also measure the static base load power at the start of each experiment to get $\pi_0$. We repeat this for $96$ power modes uniformly sampled from the $18k$ possible modes to get their specific roofline equations. For the energy roofline, we calculate energy using the empirically measured power and time and then perform a regression fit to estimate $\epsilon_{\text{flop}}$ and $\epsilon_{\text{mop}}$. 
This helps us theoretically understand the effect of static power on energy balance and energy efficiency, and model the energy efficiency in terms of the time roofline. This also offers an analytical foundation to explain our empirical results and trends.

\vspace{-0.1in}
\section{Analytical Modeling of DNN Inference}\label{sec:methodology:analytical}

We build an analytical modeling tool, \textit{Pagoda}, to estimate the FLOP and MOP for DNN workloads\footnote{The modeling tool is open-sourced at \texttt{\href{https://github.com/dream-lab/pagoda}{https://github.com/dream-lab/pagoda}}}. Using these analytical estimates, we can obtain the roofline by simply measuring the runtime and power, overcoming limitations in measuring performance counters on the edge reported in PRooF~\cite{proof_icpp24}.
Pagoda takes an ONNX representation of the DNN model as input\footnote{We initially considered PyTorch models as our input but noticed issues with getting the correct sizes for grouped convolutions for \mobilenet and hence switched to ONNX.}. 
We confirm that conversion from PyTorch to ONNX format preserves the structure of the model. We decompose each model into individual \textit{layers} and \textit{operations}, and estimate the FLOP and MOP for each. AI is obtained by dividing the cumulative FLOP by the cumulative MOP for the DNN. Our tool supports a variety of layers present in CNNs, RNNs, and transformer models. Similar to PRoof, we consider only the useful FLOP for every operation and not those due to implementation overheads~\cite{proof_icpp24}. We next discuss the process to estimate the FLOP and MOP for a convolution layer, which is most common in CNNs. Other layers are similarly derived.

\subsection{Convolution Layer Estimates}
Consider a convolution layer with an input feature map of dimensions \(N \times C_\text{in} \times H_\text{in} \times W_\text{in} \), with \(C_\text{out}\) filters of dimensions \(C_\text{in} \times K \times K\) where $N$ represents the batch size, $C_{in}$ and $C_{out}$ are the number of input and output channels respectively, $H_{in}$ and $W_{in}$ the height and width of the input feature map, $K$ is the height and width of the filter. This results in an output feature map of dimensions \(N \times C_\text{out} \times H_\text{out} \times W_\text{out}\), where $H_\text{out}$ and $W_\text{out}$ are the height and width of the output feature map respectively. Its computation tasks and associated FLOP are:
\begin{enumerate}[leftmargin=*]
    \item \textbf{Convolution Operation}: For each output element, the convolution performs \(C_\text{in} \times K \times K\) multiplications and \(C_\text{in} \times K \times K - 1\) additions, with a FLOP count of $W_\text{conv} = 2 \cdot N \cdot C_\text{out} \cdot H_\text{out} \cdot W_\text{out} \cdot C_\text{in} \cdot K \cdot K$.
    \item \textbf{Bias Addition}: Adding the bias for each output channel, its contribution to FLOP is $W_\text{bias} = N \cdot C_\text{out} \cdot H_\text{out} \cdot W_\text{out}$.
    \item \textbf{Activation Function}: Applying the activation function element-wise on the output feature map adds an additional FLOP of \( W_\text{act} = N \cdot C_\text{out} \cdot H_\text{out} \cdot W_\text{out}  \cdot c_\sigma \), where $c_\sigma$ is the FLOP for the activation function $\sigma$.
\end{enumerate}

Hence, the total FLOP for one convolution layer is:
\begin{align*}
    W_\text{conv, infer} &= W_\text{conv} + W_\text{bias} + W_{act} \nonumber \\
                         &= N \cdot C_\text{out} \cdot H_\text{out} \cdot W_\text{out} \cdot (2 \cdot C_\text{in} \cdot K \cdot K  + 1
                         +  c_\sigma)
    \label{eq:flops_conv_fwd}
\end{align*}

Similarly, the MOP for one convolution layer consists of:
\begin{enumerate}[leftmargin=*]
    \item Reading the input feature map, \( Q_\text{input}  = N \cdot C_\text{in} \cdot H_\text{in} \cdot W_\text{in} \)
    \item Reading the weights of the filters, \( Q_\text{weight} = C_\text{out} \cdot C_\text{in} \cdot K \cdot K\)
    \item Reading the biases of the filters, \( Q_\text{bias} = C_\text{out} \)
    \item Writing the output feature map, \( Q_\text{output} = N \cdot C_\text{out} \cdot H_\text{out} \cdot W_\text{out} \)
\end{enumerate}
Hence, the total MOP in bytes for one convolution layer is:
\begin{align*}
    Q_\text{conv, infer} &= Q_\text{input} + Q_\text{weight} + Q_\text{bias} + Q_\text{output} \nonumber \\
                         &= D \cdot ( N \cdot C_\text{in} \cdot H_\text{in} \cdot W_\text{in} +  C_\text{out} \cdot C_\text{in} \cdot K \cdot K + C_\text{out} \nonumber \\
                         &\quad + N  \cdot C_\text{out} \cdot H_\text{out} \cdot W_\text{out} )
\end{align*}

where $D$ is the byte-size of the datatype used for the operation.

\subsubsection{Validation using NCU}
\label{sec:ncuvalformula}
We use Nvidia's Nsight Compute (NCU)~\cite{ncu} tool to empirically confirm our analytical estimates of FLOP and MOP. Since the Jetson devices do not support measurement of all required metrics, we instead use NCU to profile these DNN workloads on an RTX 3090 GPU workstation.

The FLOP and MOP are specific to the DNN, and will not change based on the hardware used~\cite{roofline_cacm, proof_icpp24}.
We profile each workload for $20$ batches to amortize the initial warmup overheads. 
We use standard metrics and equations from~\cite{ert} to measure and calculate the FLOP and MOP using NCU.

Let the FLOP of a kernel $k$ be $F_k$, which can be expressed as the sum of FLOP from addition $F_\text{add}$, multiplications $F_\text{mul}$, and fused multiply and add $F_\text{fma}$. $F_\text{add}$, $F_\text{mul}$, and $F_\text{fma}$ are empirically measured using NCU and used to calculate the total FLOP. We also obtain the kernel's memory accesses (read and write) $M_k$ from NCU. For tensor instructions, we use a multiplying factor of 2048 on the measured FLOP in accordance with the Ampere tensor core architecture~\cite{proof_icpp24}.

The total FLOP $F_\text{total}$ and total MOP $M_\text{total}$ for the DNN are:

\begin{align*}
    F_k =& ~2 \cdot F_\text{fma} + F_\text{mul} + F_\text{add}\\
    F_\text{total} = \sum_{k \in \text{Kernels}} F_k 
\qquad & \qquad
    M_\text{total} = \sum_{k \in \text{Kernels}} M_k
\end{align*}

\section{Results and Analysis from DNN Inferencing}
\label{sec:results}

We aim to understand the energy and performance behavior of our $5$ representative DNN inference workloads: ResNet-50, MobileNetv3, YOLO-v8n, LSTM and BERT Large (Table~\ref{tbl:modeldataset}) on the Jetson Orin AGX. This section systematically analyzes these workloads by using the time and energy rooflines of various power modes and an analytical model of their compute and memory requirements. Our analysis follows a clear progression. First, in \S{\ref{subsec:ncu_val_w_analytical_model}}, we validate the accuracy of our analytical model for estimating FLOP and MOP against empirical measurements using NCU. Next, in \S{\ref{subsec:results_time_roof}}, we use the time roofline to analyze workload performance under the default MAXN power mode, examining the impact of key parameters like batch size and data precision. We then extend this analysis to understand how varying power modes affect the time roofline itself. Further, in \S{\ref{subsec:results_energy_roof}}, we leverage the energy roofline to examine energy efficiency under the default MAXN mode as well as other power modes for DNN inference, and explore the relationship between time and energy efficiency. Our analytical modeling approach, combined with the roofline framework, provides a structured way to analyze and explain the observed performance and power usage of DNN workloads across diverse power modes on the Jetson AGX Orin, where all low-level hardware metrics are not supported and cannot be measured using NCU. 

Our analysis offers two classes of insights. First, it helps reiterate well-known trends and quantify them on modern edge hardware e.g., the achievable peak compute performance is far below the theoretical specification and modestly lower for peak bandwidth, lowering the precision increases AI and performance, the energy needed for a memory operation is $36.6\times$ higher than for a FLOP, and increasing batch size offers diminishing returns on AI. Secondly, it highlights counter-intuitive observations, e.g., there are non-monotonic trends in balance points shifting as the GPU and memory frequencies are varied, the default power mode of MAXN is not the most energy efficient, time efficiency implies energy efficiency for all power modes but not vice versa, and the DNN's weights relative to its overall memory impacts $bs$ achieving memory and energy efficiency.

\begin{table*}[t!]

\footnotesize
\setlength{\tabcolsep}{3pt}
\renewcommand{\arraystretch}{1.2}
\caption{FLOP, MOP, and AI for Inference: Analytical and NCU.} 
\begin{tabular} {p{1.3cm}|p{0.8cm}|p{1cm}p{1cm}p{1cm}|p{1cm}p{1cm}p{1cm}|p{0.8cm}p{0.7cm}p{0.7cm}}
\hline
& & \multicolumn{3}{c|}{\textbf{\em Observed in NCU}} & \multicolumn{3}{c|}{\textbf{\em Analytical Estimate}} & \multicolumn{3}{c}{\textbf{\em \% Error}} \\
\textbf{Model Name} & \textbf{Batch Size} & \textbf{FLOP (G)} & \textbf{MOP (MB)} & \textbf{AI} & \textbf{FLOP (G)} & \textbf{MOP (MB)} & \textbf{AI} & \textbf{FLOP} & \textbf{MOP} & \textbf{AI} \\
\hline
\hline
ResNet & 64 & 579.89 & 22922.56 & 25.30 & 526.63 & 20810.81 & 25.31 & 9.18 & 9.21 & -0.04 \\
MobileNet & 64 & 46.44 & 7046.47 & 6.59 & 29.31 & 7469.59 & 3.92 & 36.89 & -6.00 & 40.52 \\
YOLO & 64 & 545.09 & 32160.31 & 16.95 & 565.75 & 33725.33 & 16.78 & -3.79 & -4.87 & 1.00 \\
LSTM & 32 & 0.80 & 116.92 & 6.84 & 0.75 & 103.69 & 7.23 & 6.25 & 8.73 & -5.70 \\
BERT & 64 & 5835.23 & 97854 & 59.63 & 5064.72 & 89999.93 & 56.27 & 13.20 & 8.03 & 5.21 \\
\hline
\end{tabular}
\label{tab:ncu_validation}
\end{table*}

\subsection{Validation of Analytical FLOP and MOP for Inferencing}
\label{subsec:ncu_val_w_analytical_model}

We validate our analytical FLOP and MOP against values observed using NCU for the DNN Inferencing workloads (Table~\ref{tab:ncu_validation}). This validates our analytical model's accuracy, demonstrating its reliability for roofline analysis on edge devices where direct hardware profiling is often infeasible.

For each workload, we pick a batch size that fully utilizes the GPU.

\claim{FLOP errors for our analytical model are within $10\%$ for $3$ out of $5$ DNNs and within $15\%$ for $4$ out of $5$ DNNs. MOP errors are within $10\%$ for all $5$ DNNs}

As shown in Table~\ref{tab:ncu_validation}, the FLOP errors from our Pagoda analytical model are within $15\%$ of the NCU observations, except for \mobilenet. 
The high FLOP error for \mobilenet is due to the implementation of pointwise convolution operations to reduce FLOPs. For $10$ out of $15$ ``bottleneck blocks'' in \mobilenet~\cite{mobilenet}, the channel sizes are very small and not in multiples of $64$. The tile sizes that are chosen by the NVIDIA compiler ($64$ or $128$) cause large implementation overheads of $38\%$, which form wasted work~\cite{proof_icpp24}. We verify this for a pointwise layer where we manually change the number of channels to be a multiple of the tile dimension, and here, our analytical FLOP estimates deviated from NCU by less than $6\%$. Omitting such errors due to implementation overheads, our roofline shows low FLOP errors even for \mobilenet. Also, all of our MOP errors are within $10\%$ of NCU observations for all DNNs. 

\claim{AI errors are less than $6\%$ for all DNNs except \mobilenet}
For \resnet and \yolo, AI errors are within $1\%$ of NCU observations, and for \lstm and \bert, within $6\%$. For \mobilenet, we see a $40\%$ error in AI, and this again is due to the implementation overhead due to inefficient tile sizes. 

\paragraph*{Discussion} Our analytical model offers a reasonable estimate of FLOP and MOP, but does not take into account implementation overheads and boundary conditions seen at runtime, as has been reported before~\cite{proof_icpp24}. While our analytical tool can generalize to new models, additional effort is required to correctly estimate FLOP and MOP based on workload nuances. As was reported above, there are overheads with small batch sizes due to tile sizes being larger than the matrix sizes leading to wasted FLOPs. Also,
a \textit{gemv} operation in \lstm was unrolled,
causing the weight matrix to be read multiple times~\cite{micro_LSTM}, requiring us to incorporate this into our memory estimates. Our analytical results are also consistent with those theoretically reported by others~\cite{proof_icpp24, paleo_ICLR}, validating our approach.

\vspace{-0.1in}
\subsection{Time Roofline}
\label{subsec:results_time_roof}
\subsubsection{Analysis for Default MAXN Power Mode}
\begin{figure}[t]  
    \centering  
    \begin{minipage}{0.48\columnwidth}
        \centering
        \includegraphics[width=\textwidth]{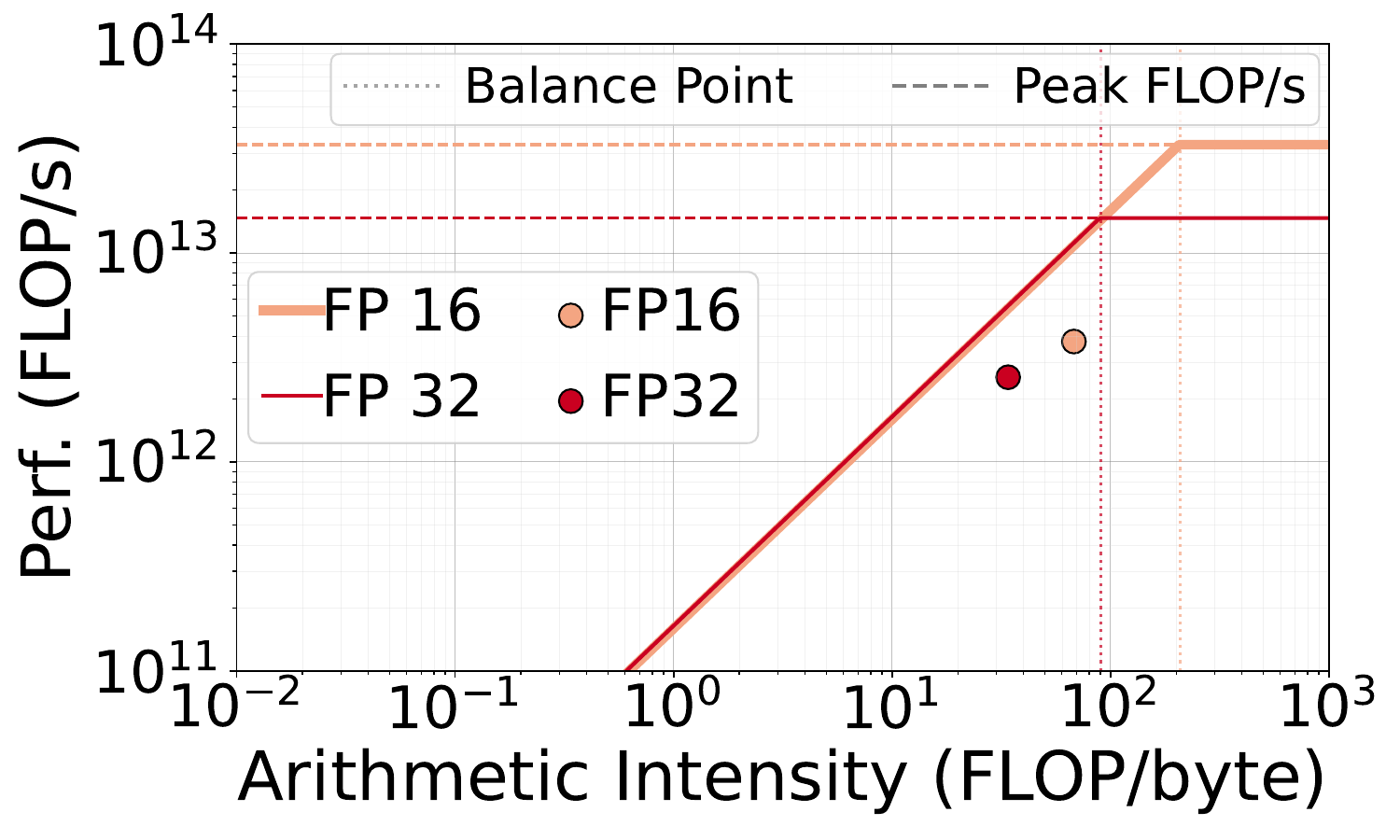}
        \caption{FP16/32 Time Roofline for Orin at MAXN}
        \label{fig:fp1632}
        \vspace{-0.1in}
    \end{minipage}  
    \hfill
    \begin{minipage}{0.48\columnwidth}
        \centering
        \includegraphics[width=\textwidth]{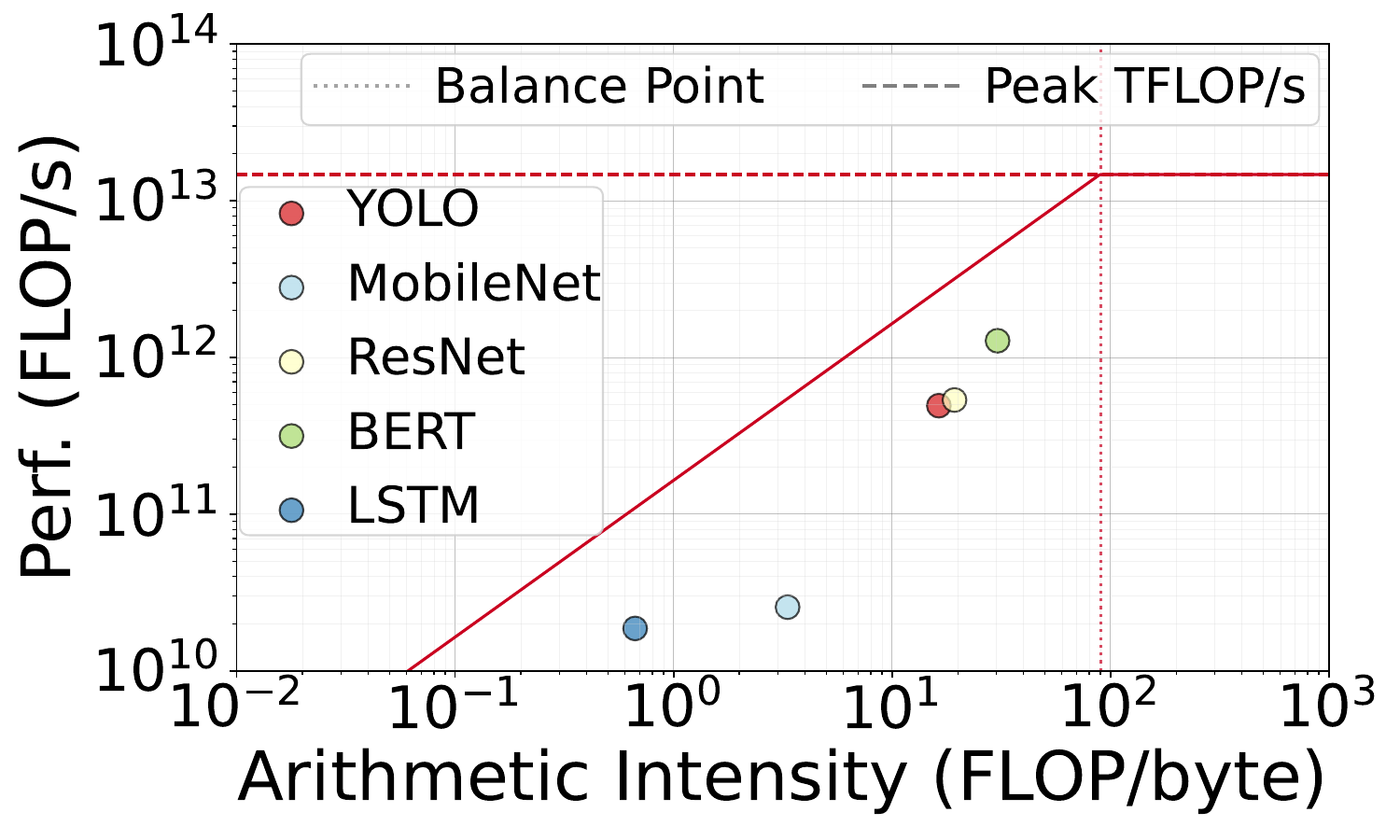}
        \caption{DNN inference on Time Roofline (MAXN, FP32)}
        \label{fig:infmodelroof}
        \vspace{-0.1in}
    \end{minipage}
\end{figure}

We use the coefficients for the Orin AGX at MAXN power mode to plot its time roofline for FP16 and FP32 precisions in Fig.~\ref{fig:fp1632}. We make several observations from its analysis.

\claim{The empirical peak performance and bandwidth for FP32 MAXN are $14.7$~TFLOP/s and $164.4$~GB/s, respectively, which are below the theoretical rating}
We observe an empirical peak compute performance of $14.7$~TFLOP/s for FP32 on the default power mode MAXN. Nvidia reports a theoretical FP32 FLOP/s for Orin AGX as $42$~TFLOP/s~\cite{Orin}, but the peak performance achieved is only $\frac{1}{3}^{rd}$ of this, at $14.7$~TFLOP/s (Fig.~\ref{fig:fp1632}). Others have reported similar empirical results~\footnote{https://forums.developer.nvidia.com/t/why-i-get-much-higher-tflops-in-orin-agx-than-what-claimed-in-the-document/311764}. We also verify peak compute performance using Nvidia's matrix multiplication benchmarks~\cite{CUDA_benchmarks}, and we get a similar $17$~TFLOP/s for FP32. Since Nvidia reports only the sparse TOP/s for INT8, we obtain the dense TOP/s for INT8 by using the rule of thumb of halving it, and further dividing it by $4$ for FP32 FLOP/s. This may give a higher than realistic theoretical estimate. 
The empirical peak memory bandwidth is $164.4$~GB/s, which is $80.3\%$ of the theoretical bandwidth of $204.8$~GB/s. We again verify this using Nvidia's memory copy benchmarks, which results in a similar bandwidth of $175$~GB/s. 

Also, as explained previously, even though we set the precision to FP32, it is automatically demoted to TF32 wherever possible to run on tensor cores. If we explicitly disable this, then we obtain the FP32 performance for just CUDA cores and not tensor cores. We report results for both of these using the Nvidia matrix multiplication benchmarks. We observe the CUDA FP32 performance as $3.49$TFLOP/s, which is about $20\%$ of the TF32 tensor core performance of $17$TFLOP/s. Since the majority of kernels from DNN workloads are executed on tensor cores, we retain the tensor core roofline through the rest of this article.

\claim{The balance point of AGX Orin on MAXN is $89.4$~FLOP/byte, which is significantly higher than older edge devices, and comparable to server-grade GPUs}
The balance point is the one at which a workload spends an equal amount of time on compute and memory operations is $89.41$~FLOP/byte for MAXN using FP32 (Fig.~\ref{fig:fp1632}).

The balance points for the older Jetsons devices, TK1 and TX1, are $18$ and $27$FLOP/byte~\cite{matei_europar}. This reflects the well-known trend of compute performance increasing much faster than memory bandwidth, which continues to hold true for edge accelerators. As a result, workloads now require substantially higher AI to fully utilize newer hardware. Interestingly, the Orin's balance point is comparable to that of workstation and server-class GPUs. E.g., the V100 and A100 have FP32 balance points of 69.5 and 101~FLOP/byte, respectively~\cite{abft_sc}. So, while modern edge devices offer greater performance, they also make it challenging for workloads to achieve balanced peak utilization.

\begin{table}[t!]
\centering
\footnotesize
\setlength{\tabcolsep}{1pt}
\renewcommand{\arraystretch}{1.2}
\caption{Estimated AI for DNN inference (bs=1).}
\vspace{-0.1in}
\begin{tabular}{l|r|r|r}
\hline
\textbf{Model Name} & \textbf{MOP (MB)} & \textbf{FLOP (G)} & \textbf{Arithmetic Intensity (AI)} \\
\hline\hline
\resnet        & 425.80   & 8.23   & 19.33\\
\yolo            & 539.35   & 8.84     & 16.39\\
\mobilenet & 138.27   & 0.46    & 3.33\\
\lstm              & 34.48    & 0.023    & 0.67\\
\bert & 2601.01 & 79.14  & 30.43\\

\hline
\end{tabular}
\label{tab:ai_bs1}
\vspace{-0.2in}
\end{table}

\subsubsection{DNN inference performance relative to the roofline}
Here, we report the performance of the DNN inference workloads in the context of the roofline and analyze their performance relative to the peak device's performance. We use the default $bs=1$ here and later study $bs$ effects. Table~\ref{tab:ai_bs1} has the estimated AI for DNN inference.

\claim{All $5$ DNN workloads are memory bound when using MAXN}
All DNNs have an AI lower than the balance point of $89.4$~FLOP/byte for the time roofline (Table~\ref{tab:ai_bs1}), i.e., they are all memory-bound. This is seen in Fig.~\ref{fig:infmodelroof}, where their performance falls below the memory roofline limit. This opens up opportunities for both system-level tuning to shift the roofline, which we explore later in this article, as well as model architecture optimizations that focus on increasing arithmetic intensity.

\claim{All DNNs achieve a performance that is well below the peak}
While the roofline gives an upper bound on the performance of DNN workloads, the actual performance achieved depends on several low-level factors such as memory access coalescing, non-FP instructions, number of FMAs, etc. \bert attains the highest performance of $1.28$~TFLOP/s, which is still less than $10\%$ of the platform peak. \mobilenet particularly is much below the roofline at a performance of $0.025$~TFLOP/s due to the pointwise convolution's inefficiency.

\claim{The performance achieved by the DNNs increases as their AI increases}

Since all five DNN models are memory-bound, the roofline model predicts their performance to be proportional to their AI. This is seen in Fig.~\ref{fig:infmodelroof}, where they lie along the diagonal, and is confirmed by our experimental performance measurements of the model inferencing. 

\subsubsection{Effect of inference batch size and precision} 
\label{bs:effect:inference}

We next study the effect of $bs$ on estimated AI and, consequently, the runtime performance.

\begin{figure}[t!]
\vspace{-0.1in}
    \centering
    \subfloat[AI vs. $bs$ for DNNs]{
        \includegraphics[width=0.50\columnwidth]{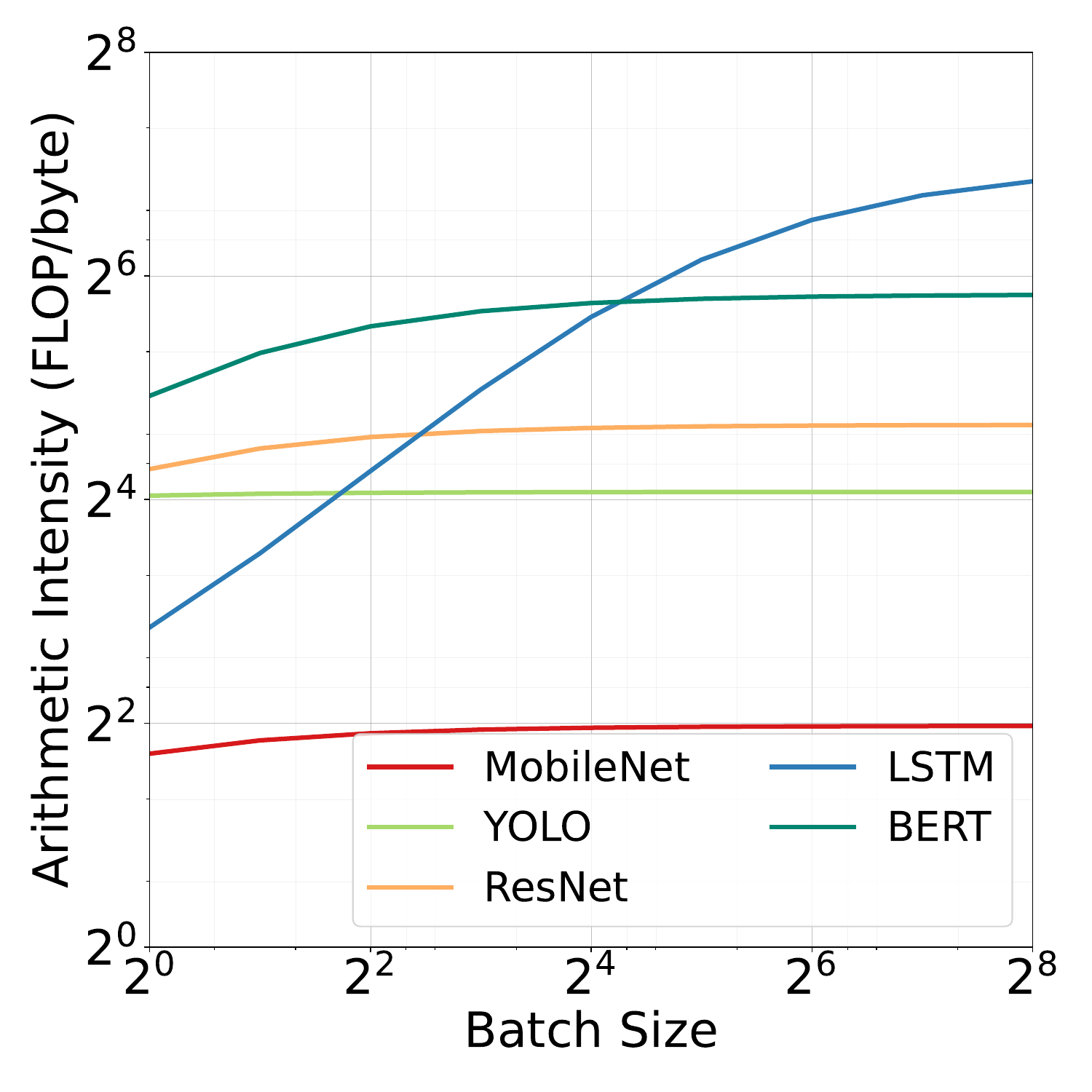}
        \label{fig:ai_bs}
    }
    \subfloat[Performance vs. AI for various $bs$]{
        \includegraphics[width=0.50\columnwidth]{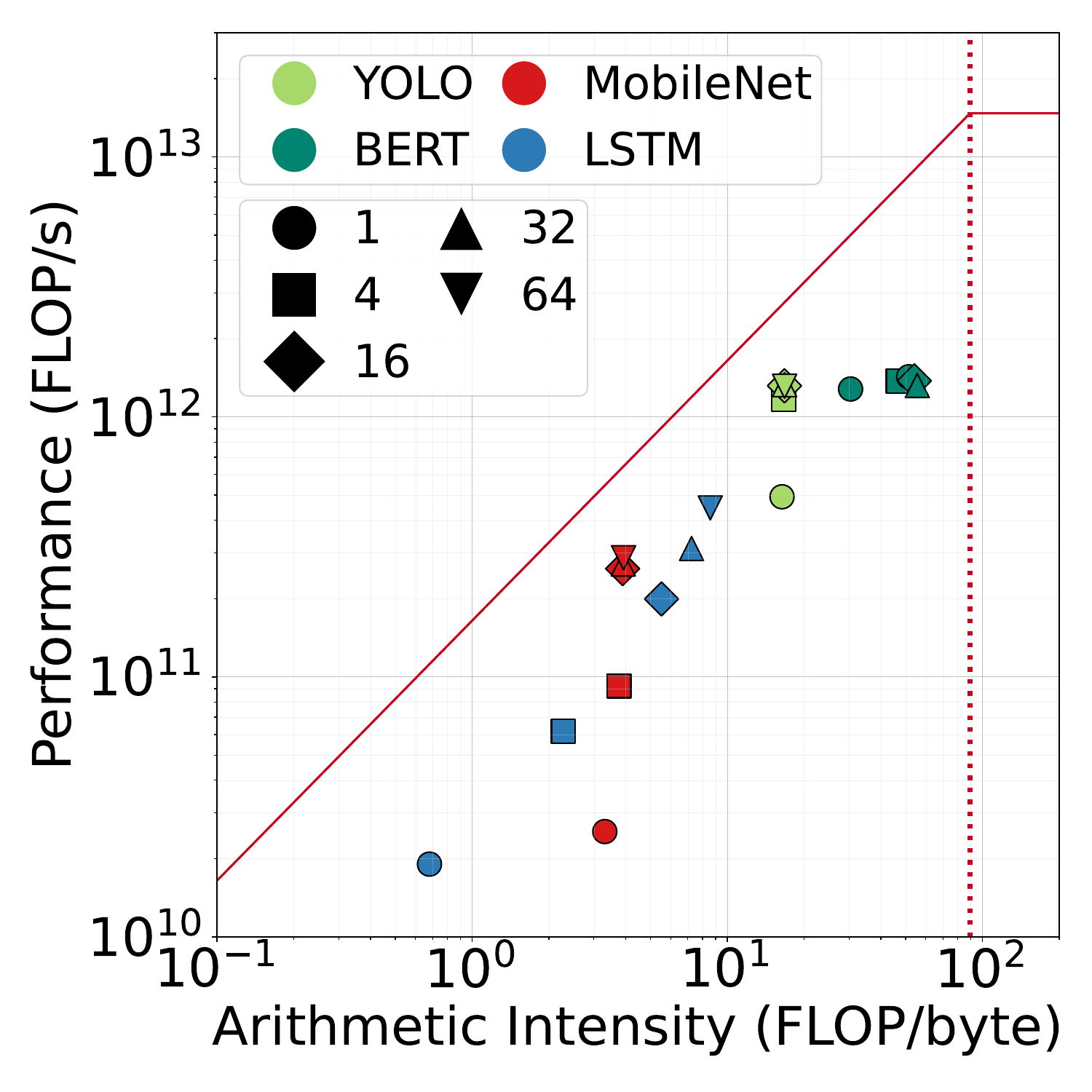}
        \label{fig:bs_performance}
    }
    \caption{Effect of inference batch size on AI and performance}
    \label{fig:bs_ai_performance}
\end{figure}

\claim{Increasing the batch size has a diminishing impact on the AI, depending on the contribution of weights to overall memory accesses}
The estimated AI for $bs=1$ for all workloads is in Table~\ref{tab:ai_bs1}. From \S~\ref{sec:methodology:analytical}, the compute (FLOP) increases linearly with batch size. However, only the input and output components of the memory scale with the batch size; the memory accesses required for reading the weights remain constant. This $bs$ growth initially causes an increase in AI. However, As $bs$ increases further, the memory sizes of the input and output outweigh the model weight, causing AI to saturate (Fig.~\ref{fig:ai_bs}).

This saturation happens at different $bs$ for the $5$ models (Fig.~\ref{fig:ai_bs}). For \yolo, at $bs=1$ the model weights contribute to just $2.3\%$
of overall memory accesses, causing the memory to linearly scale with $bs$ very early, and AI saturates early at $16.6$ for $bs=2$. For  \mobilenet, \resnet and \bert, the contributions of weights to the overall memory is $15.8, 24.0$ and $46.7\%$, and this correlates with their saturation at $bs$ of $16, 16$ and $64$. Only for \lstm, where the weights contribute to $87.4\%$ of overall memory accesses, AI increases almost linearly till $bs=1024$ before saturating. This shows that increasing batch size offers diminishing returns on AI.

\claim{All $5$ DNN workloads stay in the memory-bound region despite increasing their batch sizes}
In Fig.~\ref{fig:bs_performance}, we show the performance of all $5$ DNNs for $bs=1$ to $64$. Despite batch size increasing, none of the DNNs transition to the compute-bound region to achieve the peak performance offered by the platform. \bert at the highest batch size achieves the highest performance of $1.32$~TFLOP/s, which is still less than $10\%$ of the peak performance of $14.7$~TFLOP/s seen from the roofline.

\claim{Saturation of AI leads to saturation of workload performance, limiting benefits of $bs$ increase}
We vary $bs=1\ldots32$ for \bert and up to $64$ for the others, and measure their runtime and calculate their performance. From Fig.~\ref{fig:bs_performance}, the performance saturation corresponds to the saturation in AI for the DNNs, which is in accordance with the roofline model. E.g., for LSTM, AI linearly increases and so does performance from $0.019$ TFLOP/s at $bs=1$ to $0.45$ TFLOP/s at $bs=64$. However, for \yolo, beyond $bs=2$, the AI saturates, leading to a performance saturation as well. 

Hence, increasing batch size to increase performance only works to a certain extent, and depends on the contribution of the model weights to overall memory accesses.

\begin{figure}[t]  
    \centering  
    \begin{minipage}{0.45\columnwidth}
        \centering
        \includegraphics[width=\textwidth]{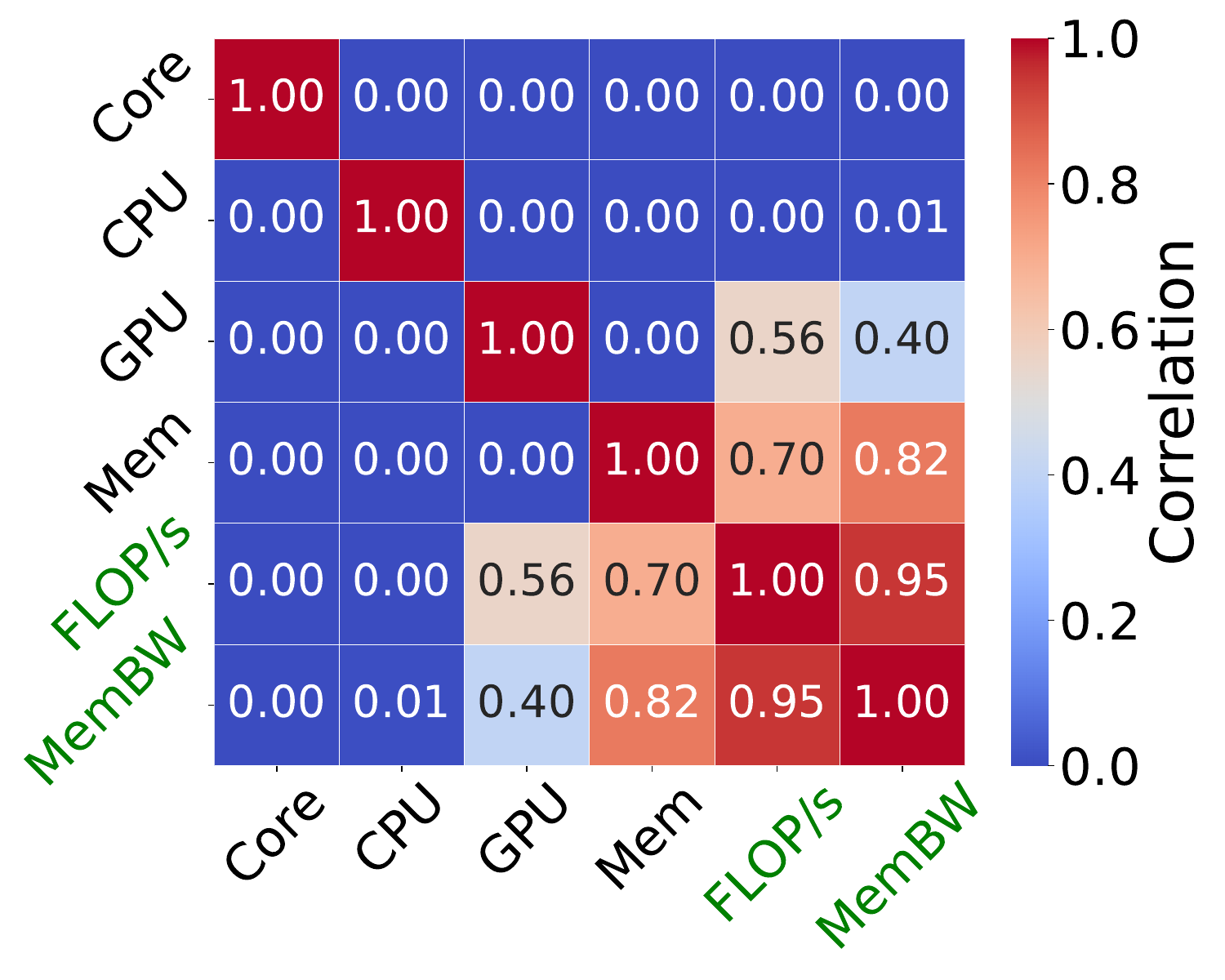}
        \vspace{-0.15in}
        \caption{Correlation Matrix: Time Roofline}
        \label{fig:corrmat}
        \vspace{-0.1in}
    \end{minipage}%
    \hfill
    \begin{minipage}{0.45\columnwidth}
        \centering
        \includegraphics[width=\textwidth]{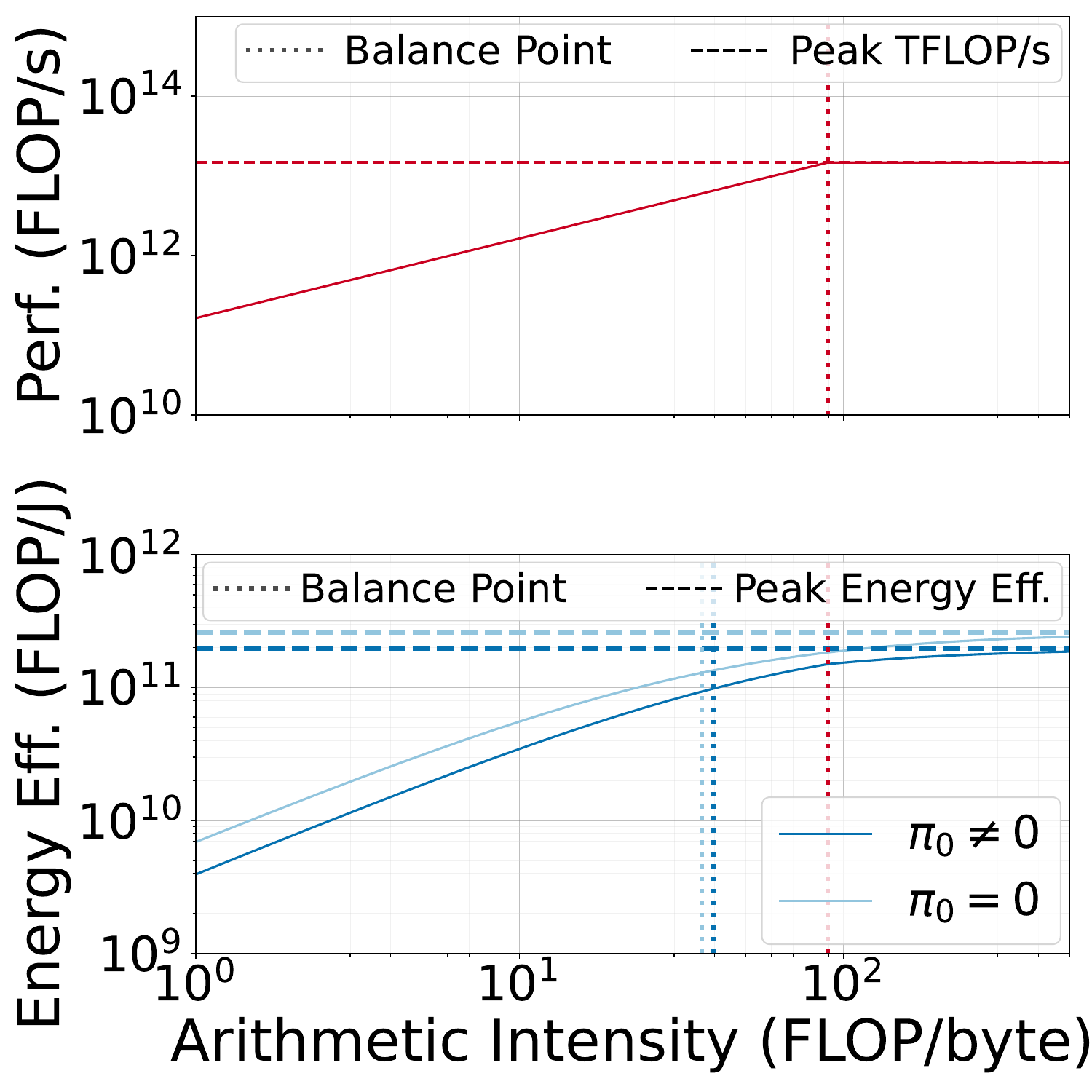}
        \vspace{-0.2in}
        \caption{Energy and Time Roofline (MAXN, FP32)}
        \label{fig:energy_roofline}
        \vspace{-0.1in}
    \end{minipage}
\end{figure}

\claim{The FP16 roofline has a higher peak performance and similar bandwidth as compared to FP32, causing the balance point to shift right}
From Fig.~\ref{fig:fp1632}, the empirical peak FLOP/s increases from $14.7$ to $33.0$~TFLOP/s, which is a $2.24\times$ increase as is expected when going from FP32 to FP16. The empirical peak bandwidth just slightly decreases from $164.4$~GB/s to $159.7$~GB/s. This causes the balance point to shift right from $89.4$~FLOP/byte for FP32 to $208.4$~FLOP/byte for FP16.

\claim{Lowering precision increases AI and performance but does not shift a workload from the memory-bound to the compute-bound region}
In going from FP32 to FP16, the FLOP remains the same, but the memory accesses halve, because each memory access is now $2$~bytes instead of $4$, causing the AI to double. However, since the balance point has shifted right by a similar amount, workloads with an AI below the balance point at a higher precision will still remain in the memory-bound region but at a higher intensity. Our empirical observations support this. In Fig.~\ref{fig:fp1632}, the balance point shifts from $89.4$~FLOP/byte for FP32 to $208.4$~FLOP/byte for FP16, and the performance of \resnet improves from $2.54$~TFLOP/s to $3.77$~TFLOP/s, but still stays in the memory-bound region even for FP16.

\subsubsection{Effect of power modes on roofline}
We obtain the time roofline coefficients for $96$ diverse power modes ($2$ core counts, $4$ CPU frequencies, $4$ GPU frequencies and $3$ memory frequencies) and analyze how they affect the FLOP/s and memory bandwidth. 

\begin{figure}[t]
\centering
\subfloat[GPU Frequency Variation.]{
    \includegraphics[width=0.49\columnwidth]{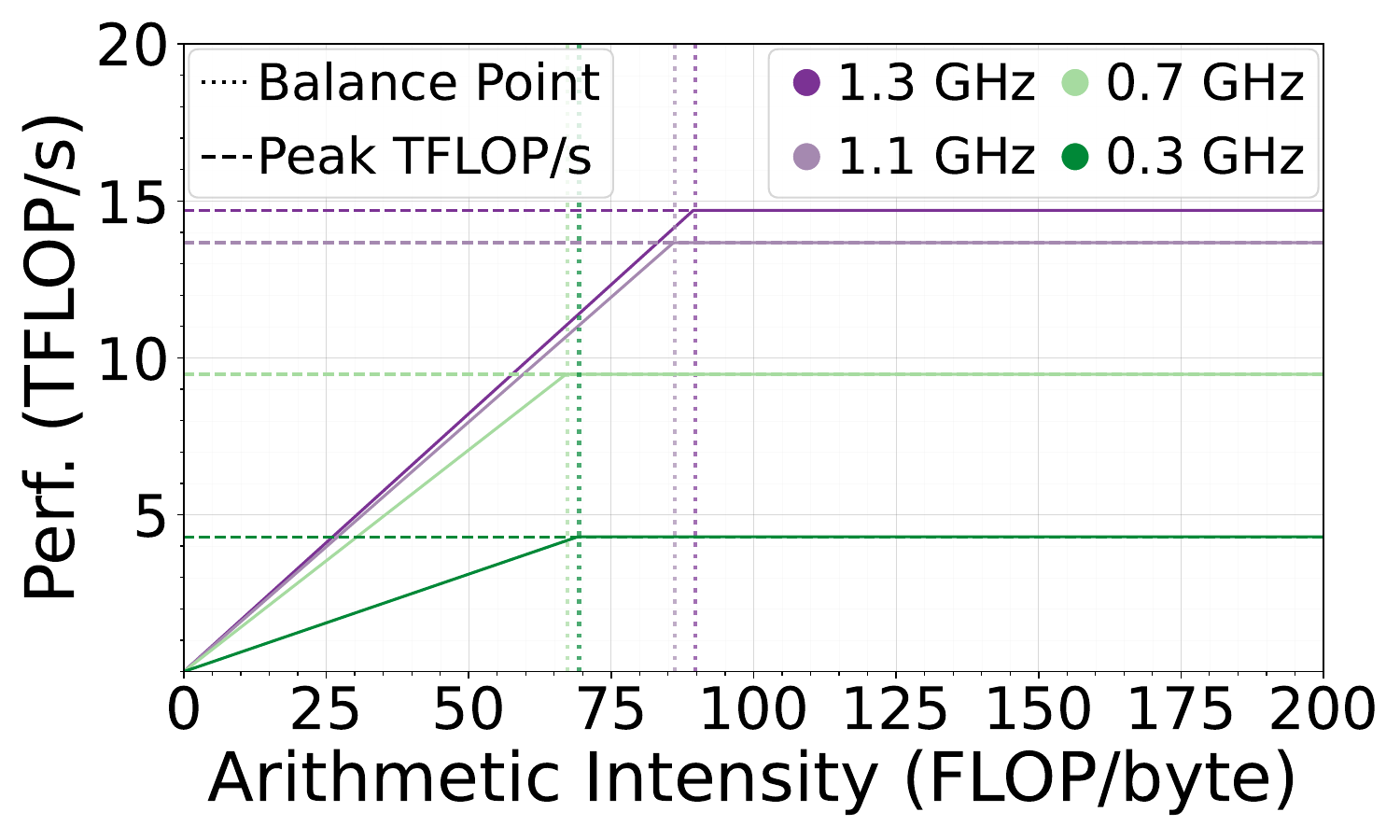}
    \label{fig:frqtime_gpu}
  }
\subfloat[Memory Frequency Variation]{
    \includegraphics[width=0.49\columnwidth]{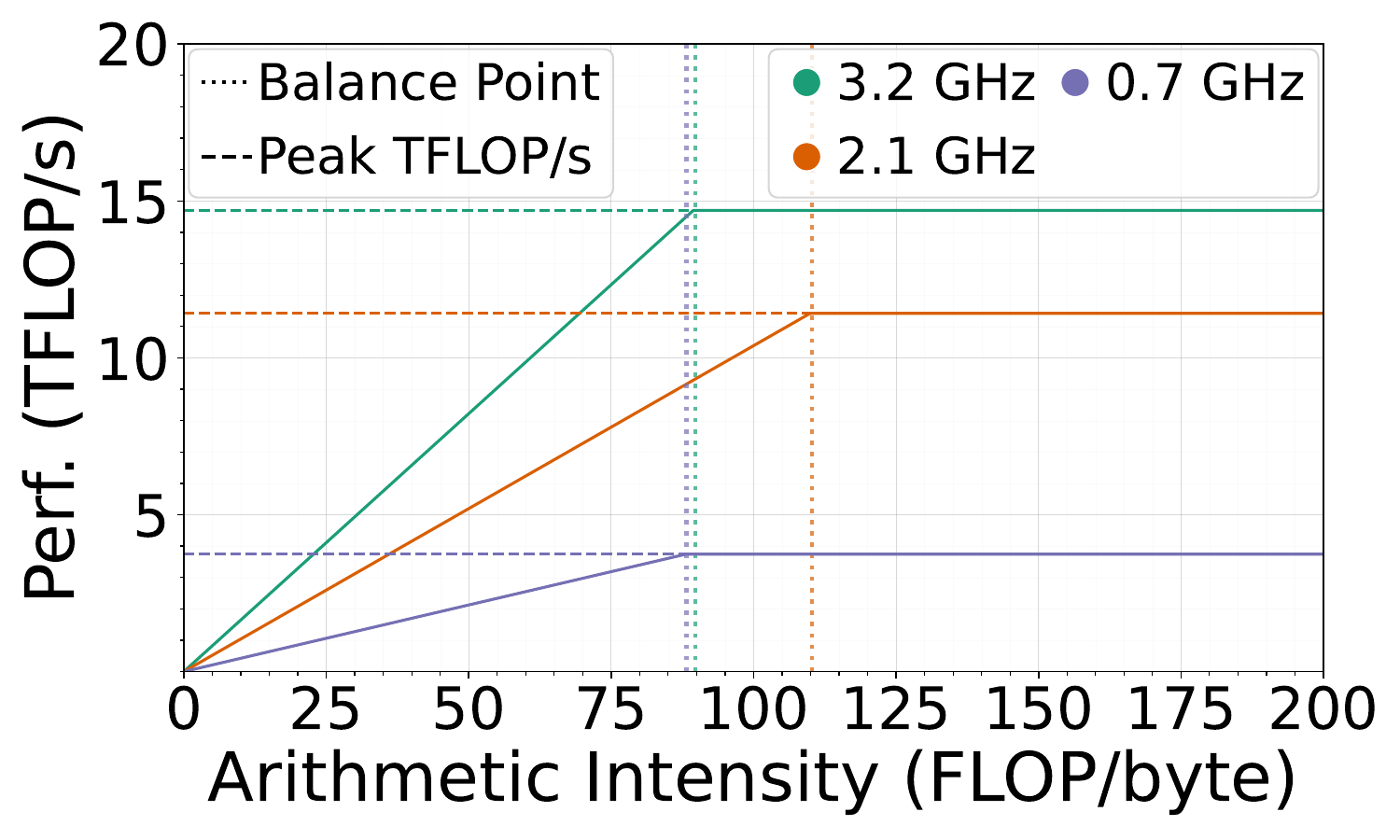}
    \label{fig:frqtime_mem}
  }

\caption{Effect of power modes on the Time Roofline}
\label{fig:frqtime_roofline}
\end{figure}

This data is used to visualize a correlation matrix (Fig.~\ref{fig:corrmat}). We plot a subset of the $96$ modes in Fig.~\ref{fig:frqtime_roofline} whose GPU or memory frequency varies while others remain constant.

\claim{CPU cores and CPU frequency have no impact on the roofline}
We experimentally observe that CPU cores and frequency do not impact the peak performance and memory bandwidth for the microbenchmarks (Fig.~\ref{fig:corrmat}), leaving the roofline unaltered. This is expected since the CPU is unrelated to the peak GPU performance and memory bandwidth. So, going forward, we only study the GPU and memory frequency changes.

\claim{Memory frequency has the highest impact on memory bandwidth and performance}

As expected, we see that changing the memory frequency for a power mode has the highest impact on memory bandwidth (Fig.~\ref{fig:corrmat}) and also on compute performance. 

The latter is due to the GPU being bottlenecked on memory operations, causing stalls, when the memory frequency is lowered. From Fig.~\ref{fig:frqtime_mem}, as memory frequency is lowered by $34.4\%$, from $3.2$GHz to $2.1$GHz, the memory bandwidth also decreases by $36.8\%$, from $164.4$~GB/s to $103.9$~GB/s. The performance drop is lesser at $\approx 22.4\%$, from $14.7$~TFLOP/s to $11.4$~TFLOP/s.

\claim{GPU frequency has the second highest impact on FLOP/s and memory bandwidth}
GPU frequency also impacts both compute performance and memory bandwidth. While the performance drop due to a lower GPU frequency is intuitive, the reduction in memory bandwidth is due to GPU being busy with compute operations at lower GPU frequencies, and under-utilizing the available memory bandwidth.

From Fig.~\ref{fig:frqtime_gpu}, when the GPU frequency reduces from $1.3$GHz to $1.1$GHz ($15.4\%$ drop), the performance decreases from $14.7$~TFLOP/s to $13.7~$TFLOP/s ($7\%$ drop), and the memory bandwidth decreases slightly by $3.2\%$. But when the GPU frequency is decreased from $1.1$GHz to $0.7$GHz ($36.4\%$ drop), the performance decreases from $13.7$~TFLOP/s to $9.5$~TFLOP/s ($30.7\%$ drop) while the bandwidth decreases from $159.1$~GB/s to $141.3$~GB/s ($11.2\%$ drop).

\claim{The balance points do not shift monotonically with variation in GPU and memory frequencies}
As shown in Fig.~\ref{fig:frqtime_gpu},  when the GPU frequency reduces from $1.3$GHz to $1.1$GHz to $0.7$GHz, the time balance point shifts left from $89.4$ to $85.9$ and further to $67.0$~FLOP/byte. However, when GPU frequency is further lowered to $0.3$GHz, the balance point unexpectedly shifts right to $69.1$~FLOP/byte. This reversal occurs because at extremely low GPU frequencies, the GPU compute becomes the bottleneck and severely under-utilizes the available memory bandwidth, flattening both compute and memory roofs. This causes the memory-bound region to dominate for a wider range of AI. Similarly, in Fig.~\ref{fig:frqtime_mem}, when the memory frequency is lowered from $3.2$GHz to $2.1$GHz, the balance point shifts right from $89.4$ to $109.9$~FLOP/byte. However, it is further lowered to $0.7$GHz, the balance point shifts left from $109.9$ to $88.1$~FLOP/byte. This reversal is because the GPU becomes bottlenecked on memory operations, effectively flattening both the roofs and allowing compute to become dominant again for lower AI. It is important to consider such non-monotonic trends when shifting rooflines for performance optimization under power constraints.

\claim{Although the power modes exhibit a wide range of frequencies, the time balance points fall in a limited range of $66.7$ to $113.3$~FLOP/byte}
While GPU and memory frequencies vary by an order of magnitude across the $96$ power modes, the variation in time balance points is curtailed. Interestingly, the peak FLOP/s across these power modes ranges from $3.4$~TFLOP/s to $14.8$~TFLOP/s, which is a change of $4.37\times$, and similarly, the memory bandwidth ranges from $35.9$~GB/s to $164.4$~GB/s, a $4.58\times$ change. However, since the time balance is a ratio of performance to bandwidth, when both are high or both are low, the ratio remains similar and this leads to the narrow range. 

\subsection{Energy roofline}
\label{subsec:results_energy_roof}
\subsubsection{Default power mode MAXN}
We first analyze the energy roofine using the MAXN power mode with peak performance.

\claim{The energy balance point (when including static power) is to the left of the time balance point}
We plot the energy and time rooflines along an aligned X axis of AI in Fig.~\ref{fig:energy_roofline}. The energy roofline with static power is in dark blue, and the energy balance point (dotted dark blue lines) is at $39.8$~FLOP/byte, which is to the left of the time balance ($89.4$~FLOP/byte). 

This implies that a workload that is compute-bound in time (time efficient) will also be compute-bound in energy (energy efficient), but the reverse is not true. 

\claim{The energy balance shifts to the left when static power is disregarded}
We also plot the energy efficiency for a hypothetical system with no base load static power consumption in Fig.~\ref{fig:energy_roofline} in light blue and its balance point (dotted light blue line). The balance point now shifts to $36.5$~FLOP/byte, which is to the left of the energy balance point when including static power. This means that workloads with a lower AI can also be energy efficient if there is no static power.

\claim{Time efficiency implies energy efficiency, regardless of static power}
Since both energy balance points are to the left of the time balance point, this means that optimizing for time automatically optimizes for energy. This is the race-to-halt situation that is commonly encountered in modern systems, where we complete a task as fast as possible and then sleep~\cite{racetohalt}. Interestingly, on MAXN, the race-to-halt is not \textit{because} of the static power's contribution to energy, unlike other systems described previously~\cite{roofline_energy_ipdps}, but \textit{regardless} of static power. In other words, even if the system designers could come up with a modified version of this edge device that did not consume any base load, race-to-halt would still hold.

\claim{The contribution of static power lowers the peak energy efficiency and also the overall energy efficiency}
The peak energy efficiency without static power is $0.26$~TFLOP/J, and it reduces to $0.196$~TFLOP/J when static power is considered. The overall energy roofline also shifts lower, indicating a lower energy efficiency throughout all, as shown in Fig.~\ref{fig:energy_roofline}. The static power is around $17.9$W for MAXN, and the peak power can go up to $67$W, meaning that the base load can form up to $26.7\%$ of overall power draw. The reduction in peak energy efficiency due to static power is $24.6\%$.

\claim{The energy needed for a memory operation is $36.6\times$ higher than the energy needed for a FLOP}
As described earlier, we use linear regression to estimate the energy per flop as $\epsilon_\text{flop}=3.86$~pJ/FLOP, and energy per memory operation as $\epsilon_\text{mop}141.38$~pJ/byte. 

Hence, the energy cost of a memory operation is $36.6\times$ that of a FLOP, for MAXN power mode.

\begin{figure*}[t!]
    \centering
    \begin{minipage}{1\columnwidth}
        \centering
        \subfloat[GPU Frequency]{
            \includegraphics[width=0.5\columnwidth]{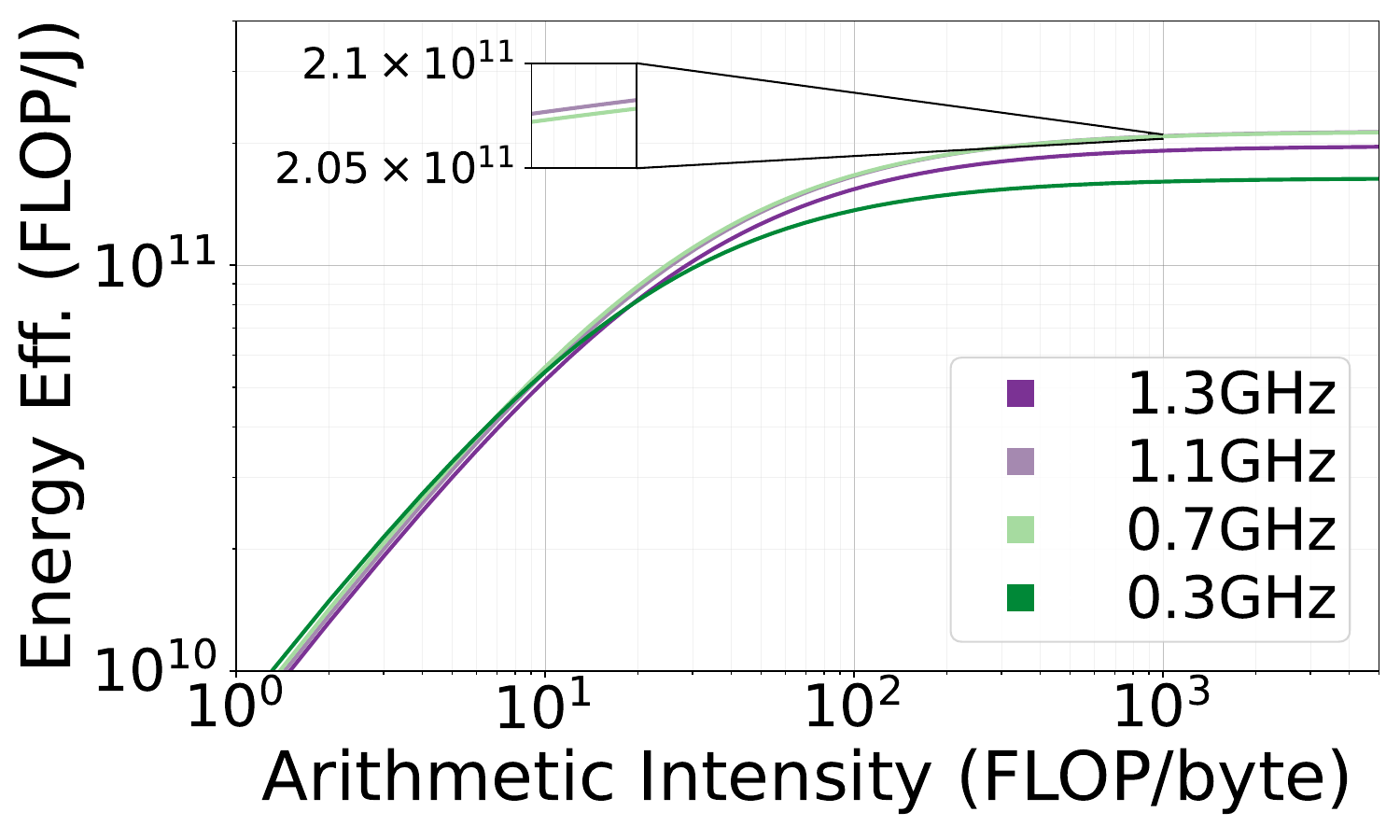}
            \label{fig:energy_GPU}
        }
        \subfloat[Memory Frequency]{
            \includegraphics[width=0.5\columnwidth]{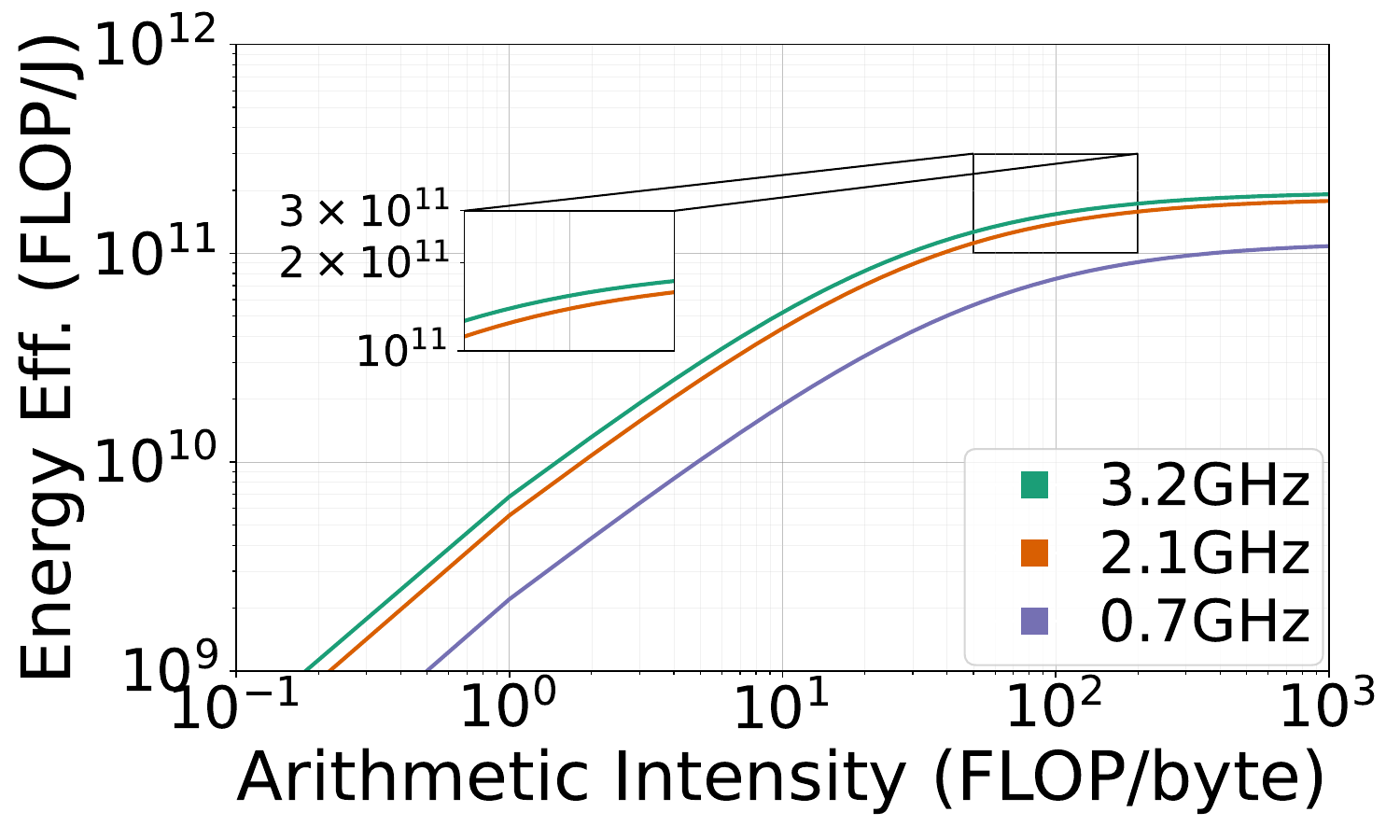}
            \label{fig:energy_mem}
        }
        \caption{Effect of power modes on the Energy Roofline}
        \label{fig:energy_frq}
    \end{minipage}
    
    \begin{minipage}{1\columnwidth}
        \centering
        \subfloat[Batch size 1]{
            \includegraphics[width=0.5\columnwidth]{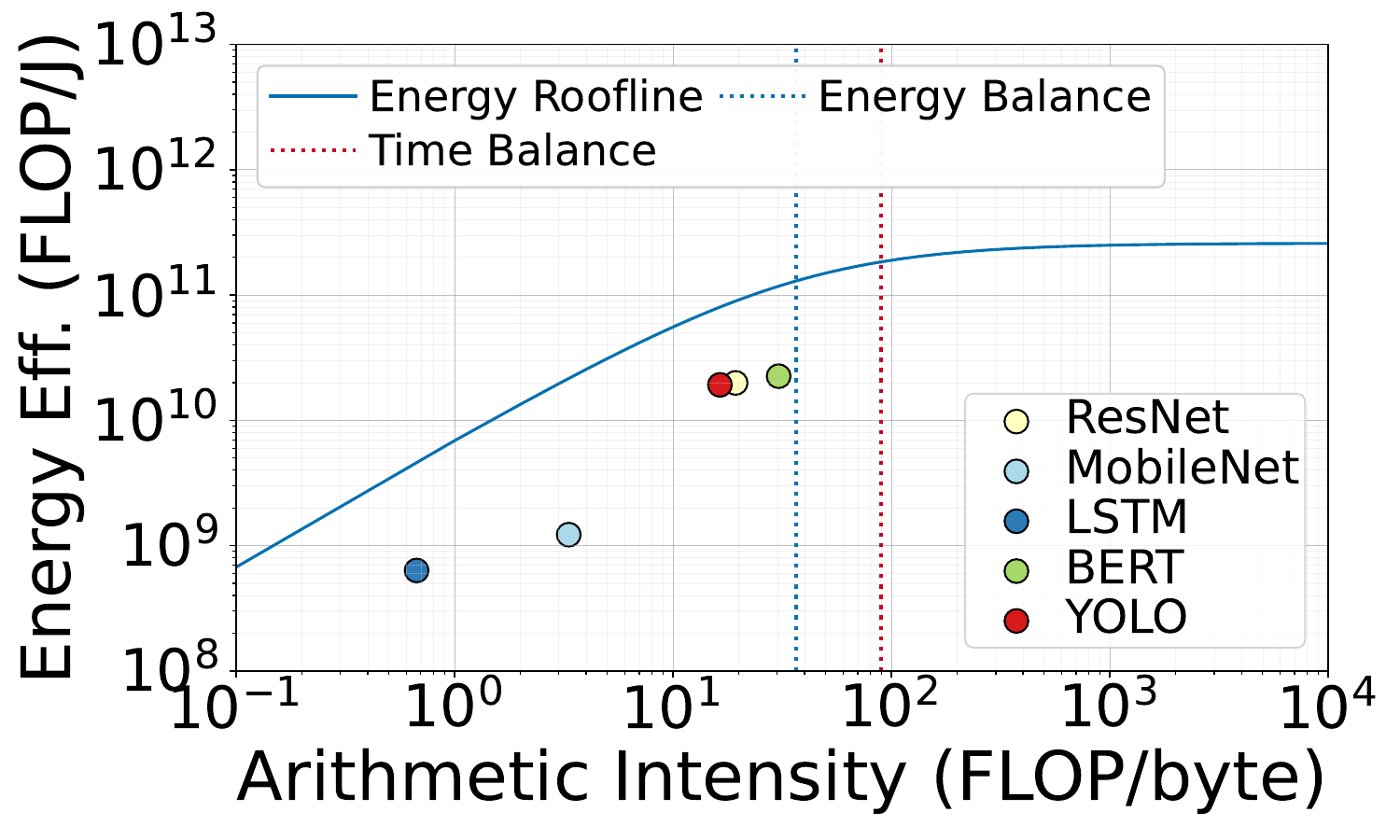}
            \label{fig:energy_in_roofline_batch1}
        }
        \subfloat[All batch sizes]{
            \includegraphics[width=0.5\columnwidth]{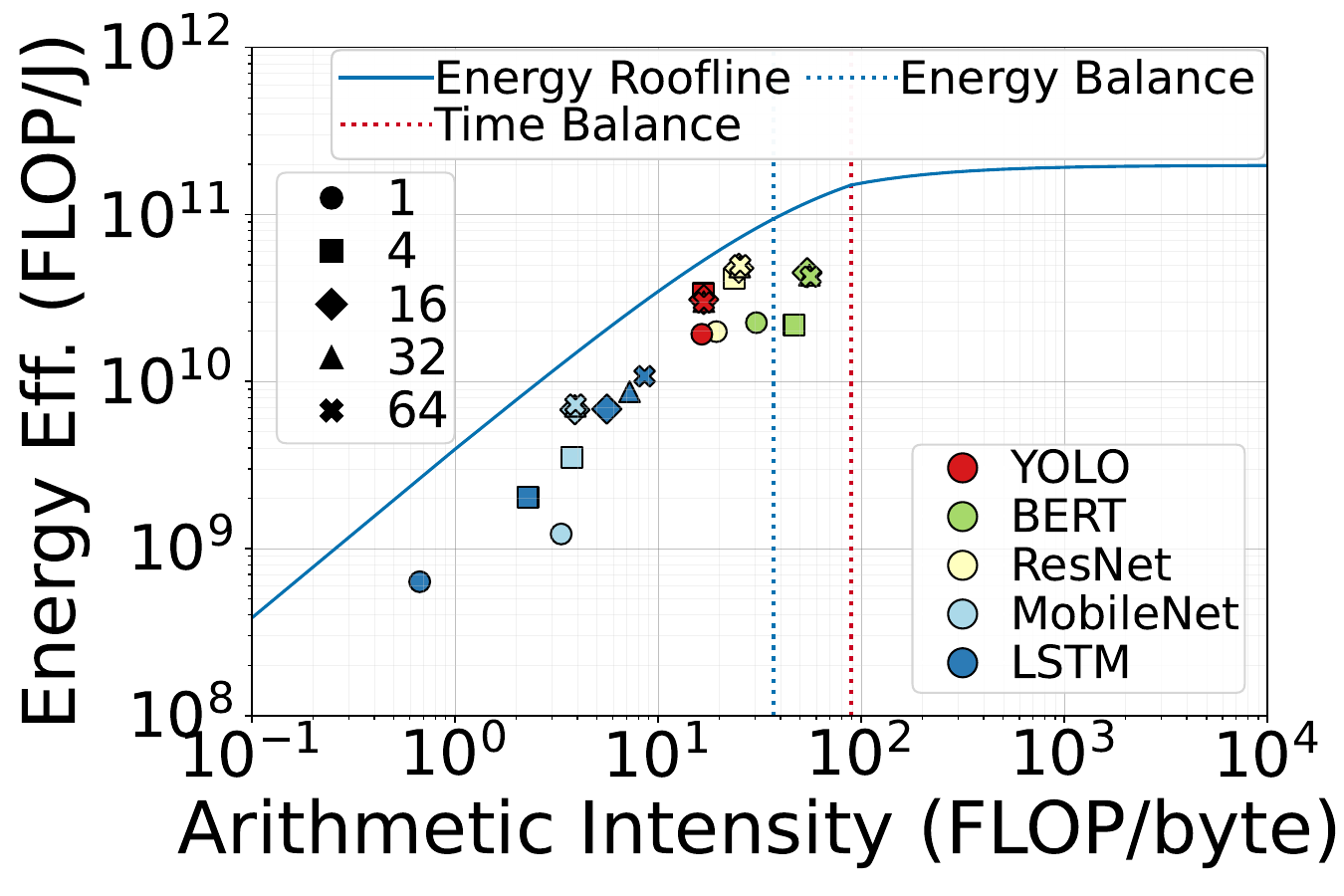}
            \label{fig:energy_in_roofline_all}
        }
        \caption{DNN Inference with Energy Roofline}
        \label{fig:energy_in_roofline}
    \end{minipage}
\end{figure*}

\begin{figure*}[t]  
    \centering
    \vspace{-0.05in}
    
    \subfloat[$\epsilon_\text{flop}$ v/s GPU Freq.]{\includegraphics[width=0.34\textwidth]{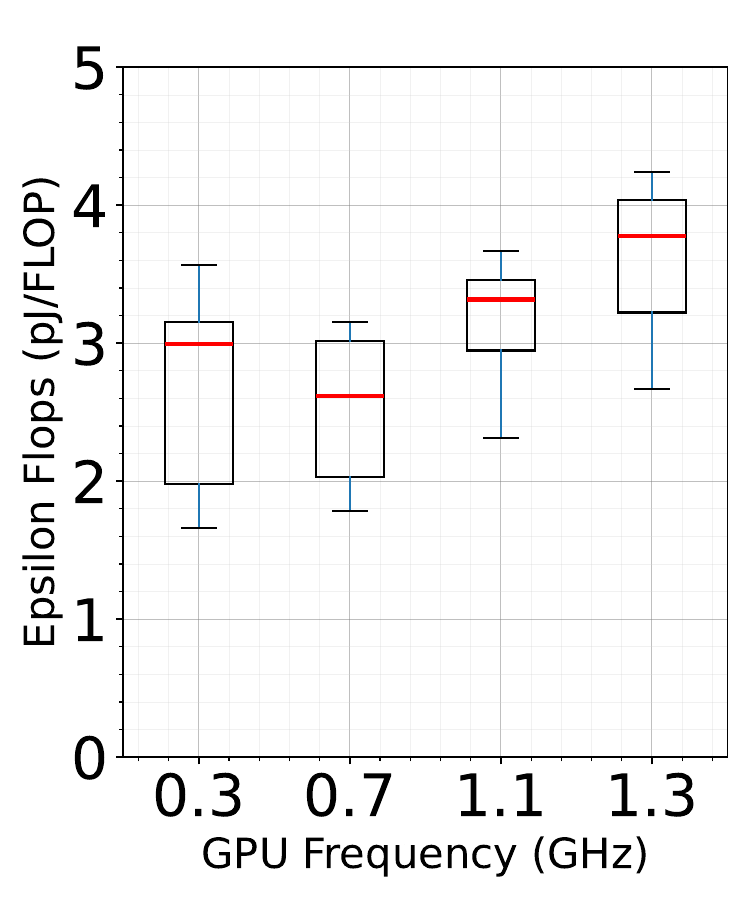}
    \label{fig:epsilonf_gpu}}
    \subfloat[$\epsilon_\text{mop}$ v/s GPU Freq.] {\includegraphics[width=0.34\textwidth]{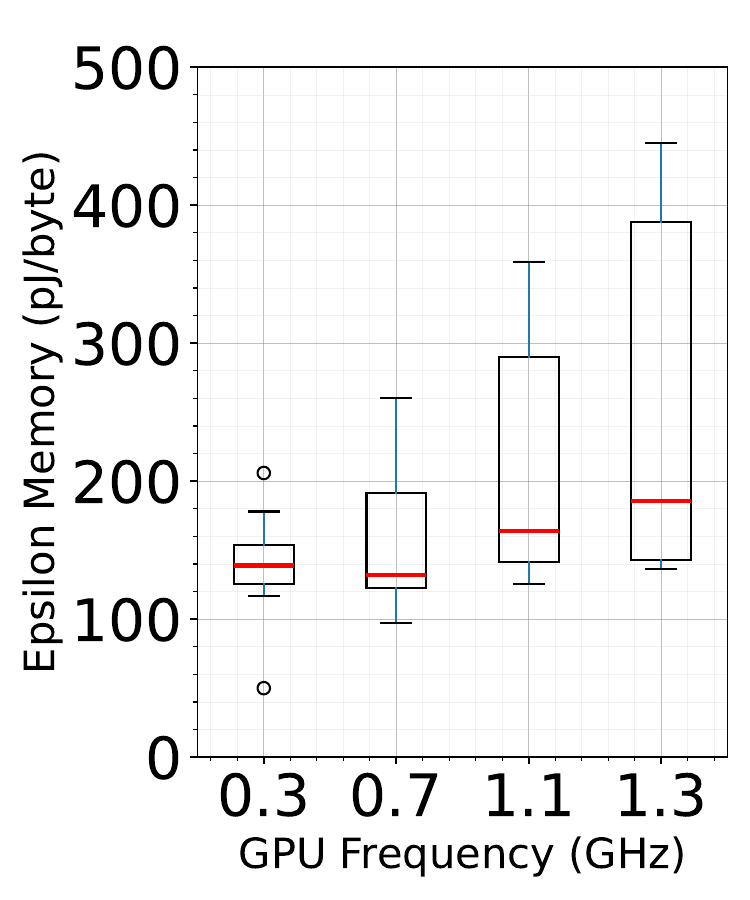}
    \label{fig:epsilonm_gpu}}
    \subfloat[$\epsilon_\text{flop}$ v/s Mem. Freq.]{\includegraphics[width=0.33\textwidth]{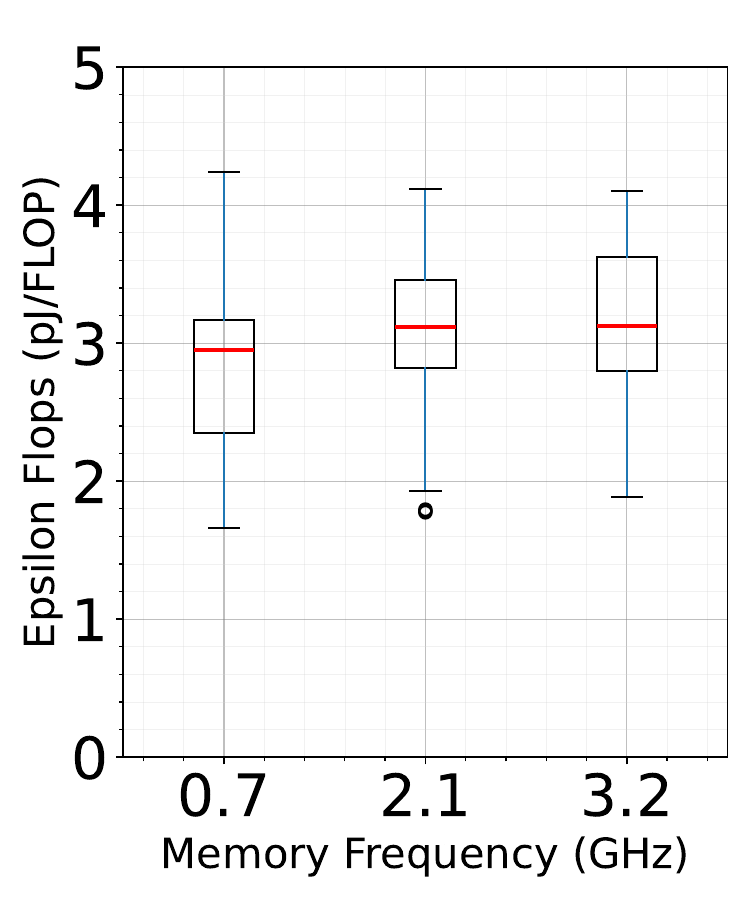}
    \label{fig:epsilonf_mem}}
    \\
    \subfloat[$\epsilon_\text{mop}$ v/s Mem. Freq.]{\includegraphics[width=0.35\textwidth]{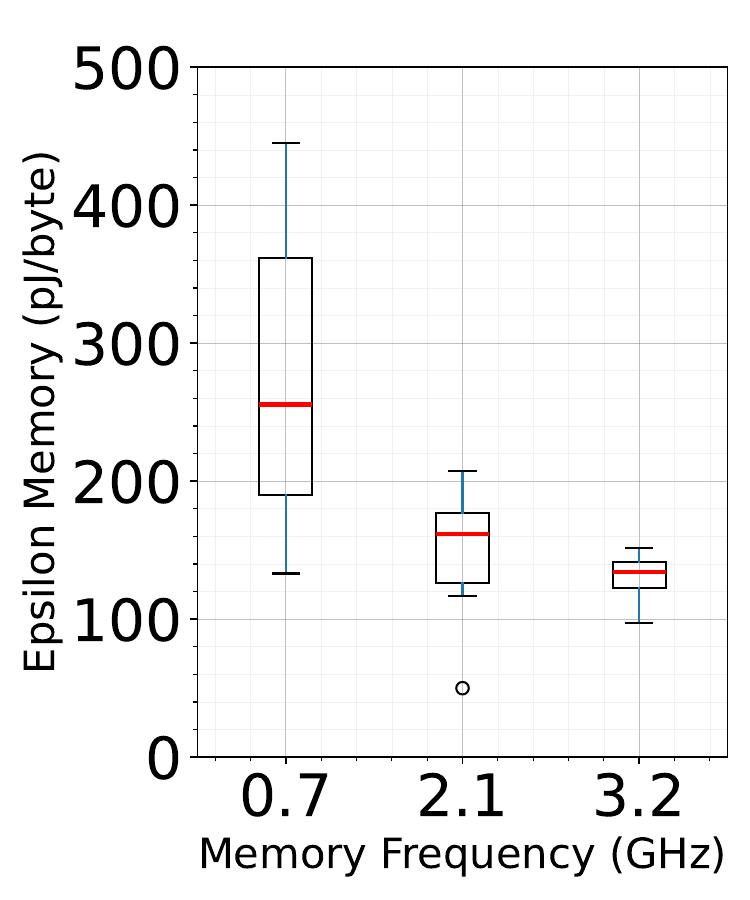}
    \label{fig:epsilonm_mem}}
    \subfloat[Power and Time]{\includegraphics[width=0.45\textwidth]{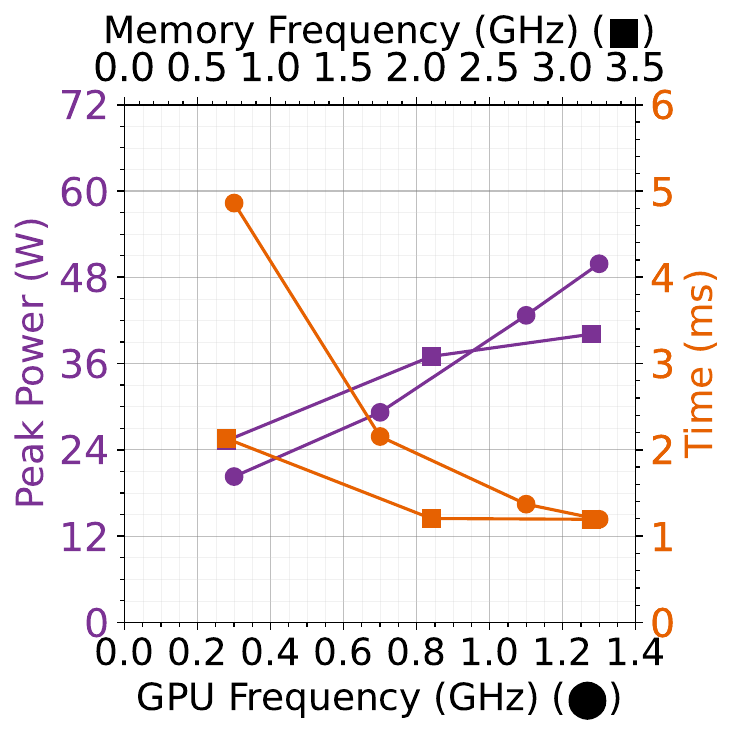}
    \label{fig:power_time}}

    \caption{Variation of $\epsilon_\text{flop}$ and $\epsilon_\text{mop}$ across power modes}
    
    \label{fig:epsilon_boxplot}
\end{figure*}

\subsubsection{Effect of power modes on the energy roofline}
In Fig.~\ref{fig:epsilon_boxplot}, we show the variation of $\epsilon_\text{flop}$ and $\epsilon_\text{mop}$ with respect to GPU frequency and Memory frequency across $96$ power modes. 

\claim{The energy per FLOP is lowest for a medium GPU frequency of $0.7$GHz, indicating that the highest GPU frequency is suboptimal for compute energy efficiency}
From Fig.~\ref{fig:epsilonf_gpu}, $0.7$GHz has the lowest median $\epsilon_\text{flop}$ across the $96$ power modes. This is due to a high time reduction when moving up to this frequency but with only a modest increase in power (Fig.~\ref{fig:power_time}), leading to an overall energy benefit.

\claim{The energy per memory operation decreases with memory frequency, and the highest memory frequency has the highest energy per MOP}
With all other dimensions at MAXN values, $\epsilon_\text{mop}$ for memory frequencies $0.7, 2.1$ and $3.2$GHz are $444.9, 174.7$ and $141.4$ respectively. Unlike GPU frequency, with memory, the time decrease and power increase comparable, leading to a monotonic behavior of energy with memory frequency. This is also seen in Fig.~\ref{fig:epsilonf_gpu} across power modes. 

\claim{GPU frequency also impacts energy for a MOP, and memory frequency also impacts energy for a FLOP}
From Figs~\ref{fig:epsilonm_gpu} and \ref{fig:epsilonf_mem}, GPU frequency has an impact on $\epsilon_\text{mop}$ and memory frequency has an impact on $\epsilon_\text{flop}$. As discussed earlier in the time roofline (\S~\ref{subsec:results_time_roof}), this is because the resources are closely connected and cause interdependencies, especially when bottlenecked.

\claim{The GPU frequency of $1.1$GHz has the highest peak energy efficiency among MAXN variants, but $0.7$GHz has the highest peak energy efficiency for most power modes}
From the inset of Fig.~\ref{fig:energy_GPU}, the GPU frequency of $1.1$GHz has the highest peak energy efficiency. This is for the power modes shown in Fig.~\ref{fig:energy_GPU}, which are MAXN and its variants with lower GPU frequencies. However, the energy efficiency of $1.1$GHz is better than that of $0.7$GHz by just $0.3\%$.
Further, looking at the median $\epsilon_\text{flop}$ across all power modes in Fig.~\ref{fig:epsilonf_gpu}, a GPU frequency of $0.7$GHz has the lowest $\epsilon_\text{flop}$. From Eqns.~\ref{eq:Final EquationA} and \ref{eq:Final EquationB}, $\epsilon_\text{flop}$ and $\epsilon_\text{mop}$ are inversely proportional to energy efficiency. In the compute-bound region, $\epsilon_\text{flop}$ has a greater impact on energy efficiency than $\epsilon_\text{mop}$. Accordingly, we see that $0.7$GHz has the highest peak energy efficiency for $22$ out of $24$ power modes; the second highest is $\approx 6.4\%$ lower than this. So, the highest GPU frequency is not the most energy-efficient.

\claim{The peak memory frequency of $3.2$GHz has the highest energy efficiency among MAXN variants, but $2.1$GHz has the highest energy efficiency for most power modes} 
From Fig.~\ref{fig:energy_mem}, the peak memory frequency of $3.2$GHz has the best energy efficiency, followed by $2.1$ and $0.7$GHz in the memory-bound and compute-bound regions, for MAXN and its variants with lower memory frequencies. However, this is not always the case. In Fig.~\ref{fig:epsilonf_mem}, the median $\epsilon_\text{flop}$ is slightly lower ($<1\%$) for memory frequency $2.1$GHz than $3.2$GHz, and this causes the energy efficiency to be slightly higher for $2.1$GHz in several cases ($18$ out of $32$). E.g., the power mode with $8$ CPU cores, $1651$MHz CPU, $727$MHz GPU and $2.1$GHz memory has a $2.4\%$ higher energy efficiency than MAXN with $3.2$GHz memory. So, the highest memory frequency is not always the most energy-efficient one.

\claim{Among the $96$ power modes, the one with the highest peak energy efficiency has a GPU frequency of $0.7$GHz and a memory frequency of $2.1$GHz}
The power mode with the highest peak energy efficiency is not MAXN, which is the default power mode with a GPU frequency of $1.3$GHz and memory frequency of $3.2$GHz. Instead, it is a lower power mode with $0.7$GHz GPU and $2.1$GHz memory. As discussed, this GPU frequency has the lowest $\epsilon_\text{flop}$ and, hence, a high energy efficiency. In the compute-bound region, $\epsilon_\text{flop}$ has a greater impact on energy efficiency than $\epsilon_\text{mop}$ due to high AI. From Fig.~\ref{fig:epsilonf_mem}, the median $\epsilon_\text{flop}$ is slightly lower ($<1\%$) for memory frequency $2.1$GHz than $3.2$GHz, causing energy efficiency to be higher for $2.1$GHz.

\claim{For all $96$ power modes, time efficiency implies energy efficiency}
For all $96$ power modes, the energy balance (with static power) falls to the left of time (not shown in Fig.), and hence time efficiency implies energy efficiency. This empirically validates the idea proposed in previous work~\cite{roofline_energy_ipdps}.

\claim{For some power modes, when static power is ignored, energy efficiency implies time efficiency}
We find that for $24$ out of $96$ power modes, the energy balance point without static power lies to the right of the time balance point (not shown in Fig.), i.e., energy efficiency implies time efficiency for these. Interestingly, $21$ out of these power modes have the lowest memory frequency of $0.7$GHz. From Fig.~\ref{fig:epsilon_boxplot}, the lowest memory frequency has the highest $\epsilon_\text{mop}$ and the lowest $\epsilon_\text{flop}$. From Eqn.~\ref{eq:energy_balance_non_zero_pi_not}, this results in the highest energy balance point, explaining the crossover.

\claim{For these power modes with a time energy balance point crossover, static power is the reason for race-to-halt to hold}
When a power mode exhibits a time-energy balance point crossover, it is compute-bound in terms of energy but memory-bound in terms of time. While this implies that reducing compute-related energy would be the most effective strategy for energy efficiency, this is not the case because of the device's static power consumption.

If system designers were able to eliminate this static power, the race-to-halt principle would no longer apply to these modes. Instead, the most energy-efficient strategy would be to operate at a lower, more balanced frequency to reduce the dynamic power consumed by the compute unit, as there would be no static power penalty for a longer runtime.

\subsubsection{Inference performance relative to the roofline}
Here, we present the inference performance of various DNN workloads relative to the energy roofline. We also investigate the effect of batch size on energy efficiency.

\claim{Workloads show energy efficiencies well below the roofline}
In Fig.~\ref{fig:energy_in_roofline}, the energy efficiencies of all $5$ DNN models are well below the energy roofline. This is due to not just inefficiencies of workloads, as seen in the time roofline (\S~\ref{subsec:results_time_roof}), but also the power consumption and how much it differs from the peak power.

\claim{All $5$ DNNs are memory-bound in energy}
All $4$ DNNs have AIs less than the energy balance ($39.8$~FLOP/byte) of MAXN and are memory-bound in energy. Therefore, they spend more energy on memory than compute, resulting in a lower energy efficiency.

\claim{\lstm has the lowest energy efficiency and \bert has the highest energy efficiency, in accordance with their AIs}
As seen in Fig.~\ref{fig:energy_in_roofline}, the energy efficiencies of various DNNs are in order of increasing AIs. Thus, \lstm has the lowest energy efficiency of $0.6$~GFLOP/J and \bert has the highest energy efficiency of $22.5$~GFLOP/J.

\claim{\bert at higher batch sizes is compute-bound in energy but memory-bound in time}
From Fig~\ref{fig:energy_in_roofline_all}, the AI of \bert beyond $bs=1$ is greater than the energy balance point but less than the time balance point. This means that \bert is simultaneously compute-bound in energy but also memory-bound in time, i.e., the majority of time spent is on memory operations, but the majority of energy spent is on compute operations.

\claim{Increasing batch size increases energy efficiency, but again has diminishing returns due to saturation in AI}
As explained in \S~\ref{bs:effect:inference}, \yolo's AI saturates very quickly while \lstm's AI increases linearly till $bs=1024$, and the same trend is reflected in the energy efficiency as well. From Fig~\ref{fig:energy_in_roofline_all}, we see that \lstm's energy efficiency increases $17\times$ from $0.6$~GFLOP/J at $bs=1$ to $10.8$~GFLOP/J at $bs=64$. \yolo's energy efficiency only modestly increases by $55.5\%$, from $19.1$~GFLOP/J at $bs=1$ to $29.7$~GFLOP/J at $bs=64$. Thus, the contribution of a DNN's weights to its overall memory decides the batch size at which AI performance and energy efficiency saturate -- higher the contribution, more the benefit from batch size increases.

\section{Analysis of DNN Training}
\label{sec:training}
We extend our analytical model to include DNN training workloads, where the key distinction over inferencing is the inclusion of the backward pass to propagate the loss. We derive for FLOP for the \textit{backward pass} of a convolution layer as:

\begin{align*}
    W_\text{backward} &= 2 \cdot N \cdot C_\text{in} \cdot H_\text{in} \cdot W_\text{in} \cdot \left( C_\text{out} \cdot K \cdot K \right) \nonumber \\
    &\quad + 2 \cdot N \cdot C_\text{out} \cdot H_\text{out} \cdot W_\text{out} \left( C_\text{in} \cdot K \cdot K \right) \quad \text{flop}
\end{align*} 

The two components are for computing the gradients with respect to the input activations and weights respectively. Similarly, memory accesses for the backward pass involve reading the output gradients, weights, input feature map and output gradient, and writing input gradients, weight and bias gradients. This is computed as:
\begin{align*}
    Q_\text{backward} &= 2 \cdot N \cdot C_\text{in} \cdot H_\text{in} \cdot W_\text{in} + 2 \cdot N \cdot C_\text{out} \cdot H_\text{out} \cdot W_\text{out} \nonumber \\
    &\quad + 2 \cdot C_\text{out} \cdot C_\text{in} \cdot K \cdot K + C_\text{out} \quad \text{mop.}
\end{align*}
This gives us:

\begin{align*}
    W_\text{train} \approx 3 \cdot W_\text{inference} \quad \text{and} \quad
    Q_\text{train} \approx 3 \cdot Q_\text{inference}
\end{align*}

For brevity, we present concise results on DNN training. All training DNNs are run with $bs=16$, as mentioned earlier.

\begin{figure}[t]
\vspace{-0.3cm}
\centering
\subfloat[Time Roofline]{
\includegraphics[width=0.5\columnwidth]{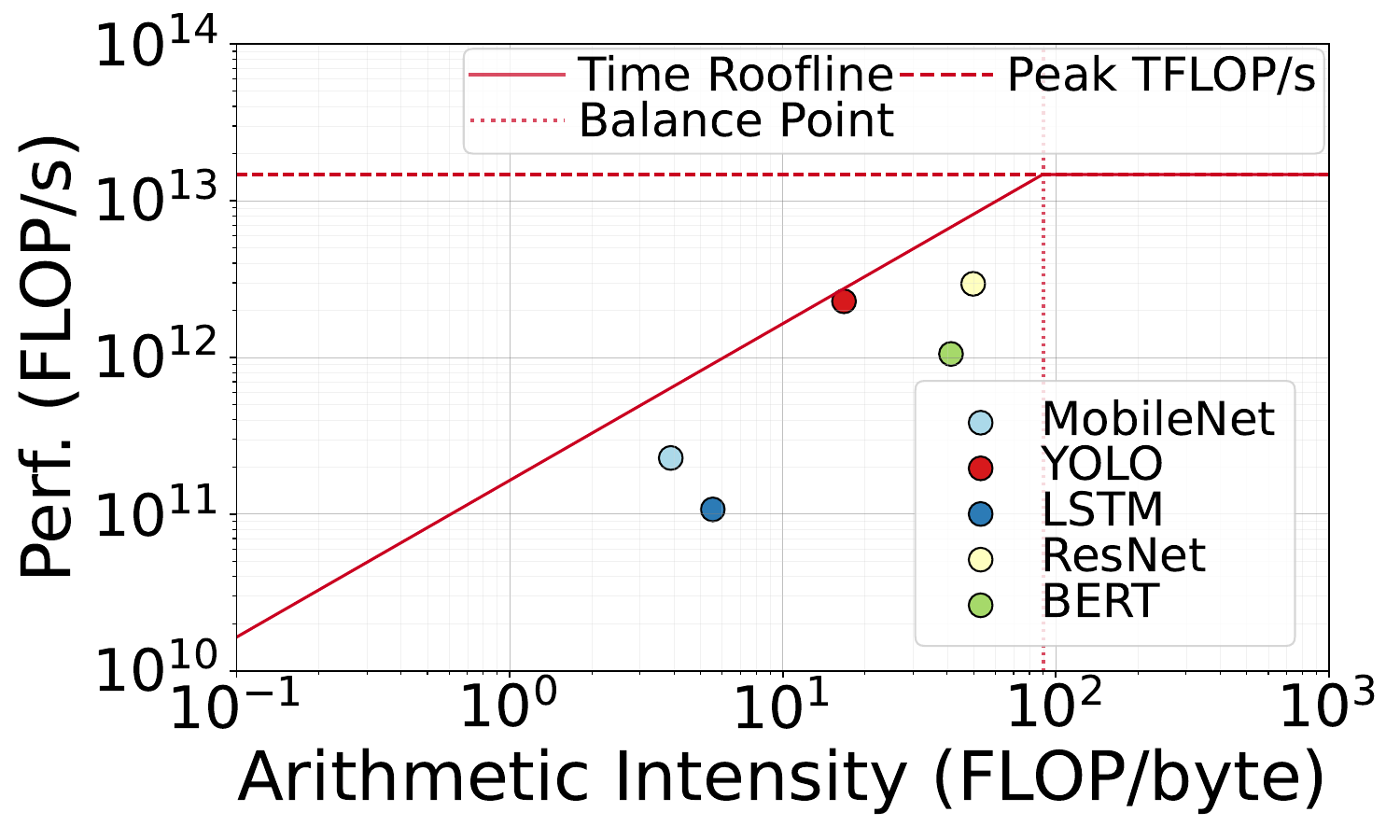}
\label{fig:time_train_roofline}
}
\subfloat[Energy Roofline]{
\includegraphics[width=0.5\columnwidth]{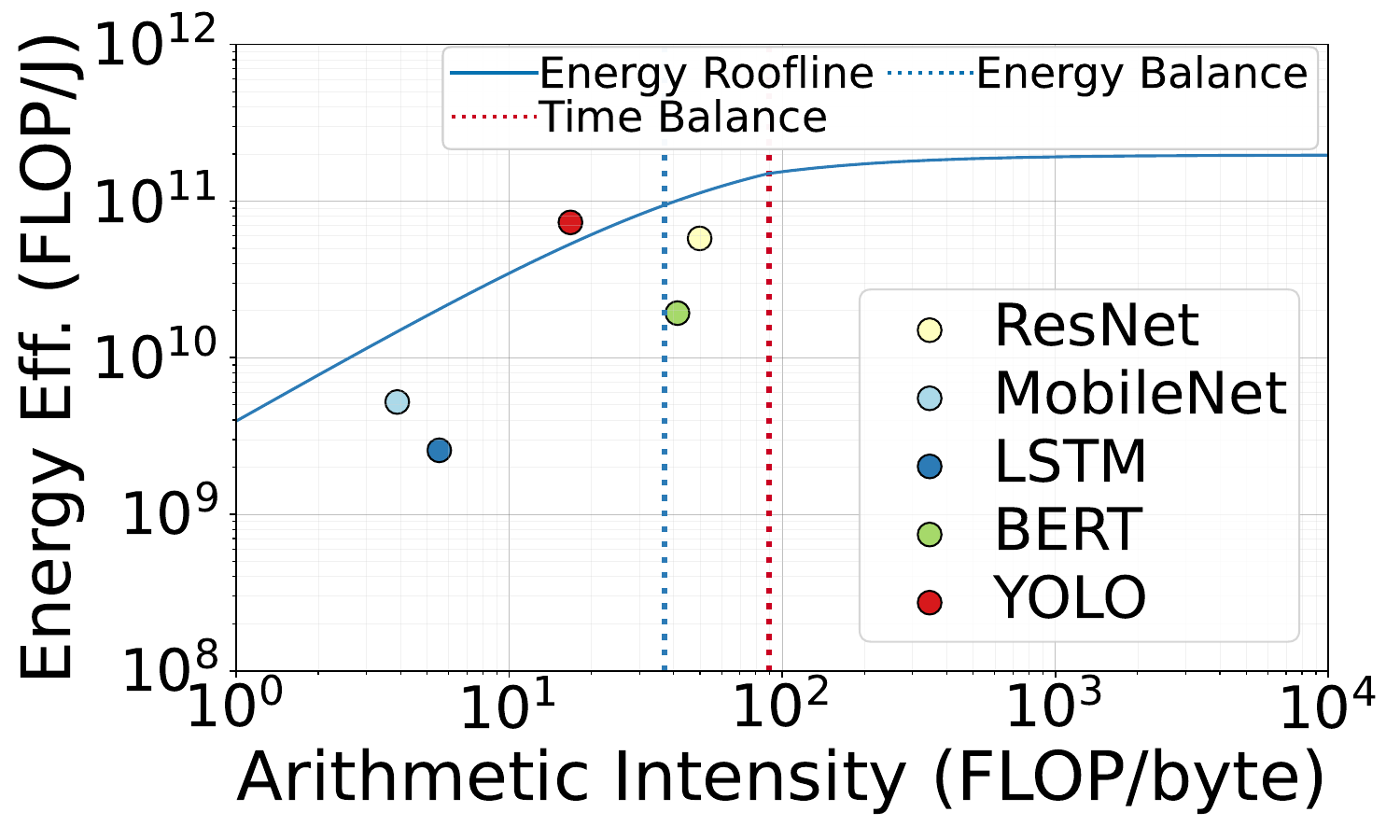}\label{fig:energy_train_roofline}}

\caption{DNN Training}
\label{fig:train_roofline}
\end{figure}

\claim{All $5$ DNNs are memory-bound in time even for training}
From Fig.~\ref{fig:time_train_roofline}, all $5$ DNNs are memory-bound and fall to the left of the time balance point. This is because both FLOP and MOP increase by $3\times$ for training relative to inference.
So, even though training performs more computation than inference, their AI does not vary for a batch size.

\claim{$3$ of $5$ DNNs are compute-bound in energy, and $2$ are simultaneously memory-bound in time}
In Fig.~\ref{fig:energy_train_roofline}, the AI of \bert and \resnet falls between the time and energy balance points, indicating that they spend most of their energy on compute but a majority of time on memory.
\mobilenet, \lstm and \yolo are memory-bound in both energy and time.

\claim{\lstm has the lowest performance and energy efficiency in training, and \yolo falls close to the roofline, exhibiting a high hardware efficiency}
\lstm has the lowest performance and energy efficiency, and the implementation overheads discussed in inference hold here as well. Surprisingly, \yolo has much better performance compared to the others and falls close to the time roofline, and slightly above the energy roofline. This requires further investigation.

\paragraph*{Discussion} While our estimates of training FLOP and memory being $3\times$ hold for convolution and most other CNN layers, further examination is required into whether this holds for RNNs, transformers and other architectures that feature different layers. If so, our analytical model may need to be enhanced to incorporate such layer-specific calculations.

\section{Using Roofline to Optimize DNN Inference}

\label{sec:casestudy}
Using insights from our DNN inference study and the roofline model, we examine if we can change the power mode to help shift the device roofline to match the DNN workload. We present early results on this optimization.

\subsection{Time roofline based optimization}
According to the time roofline, the workload is bottle-necked either on memory or compute. So, only GPU and memory frequencies are critical, and the other dimensions of CPU cores and frequency are over-provisioned. Lowering these can result in power savings without impacting performance, and lead to energy benefits. We investigate this possibility and try to estimate performance when the roofline shifts. 

\begin{figure}[t]
\centering
\subfloat[ResNet]{
    \includegraphics[width=0.49\columnwidth]{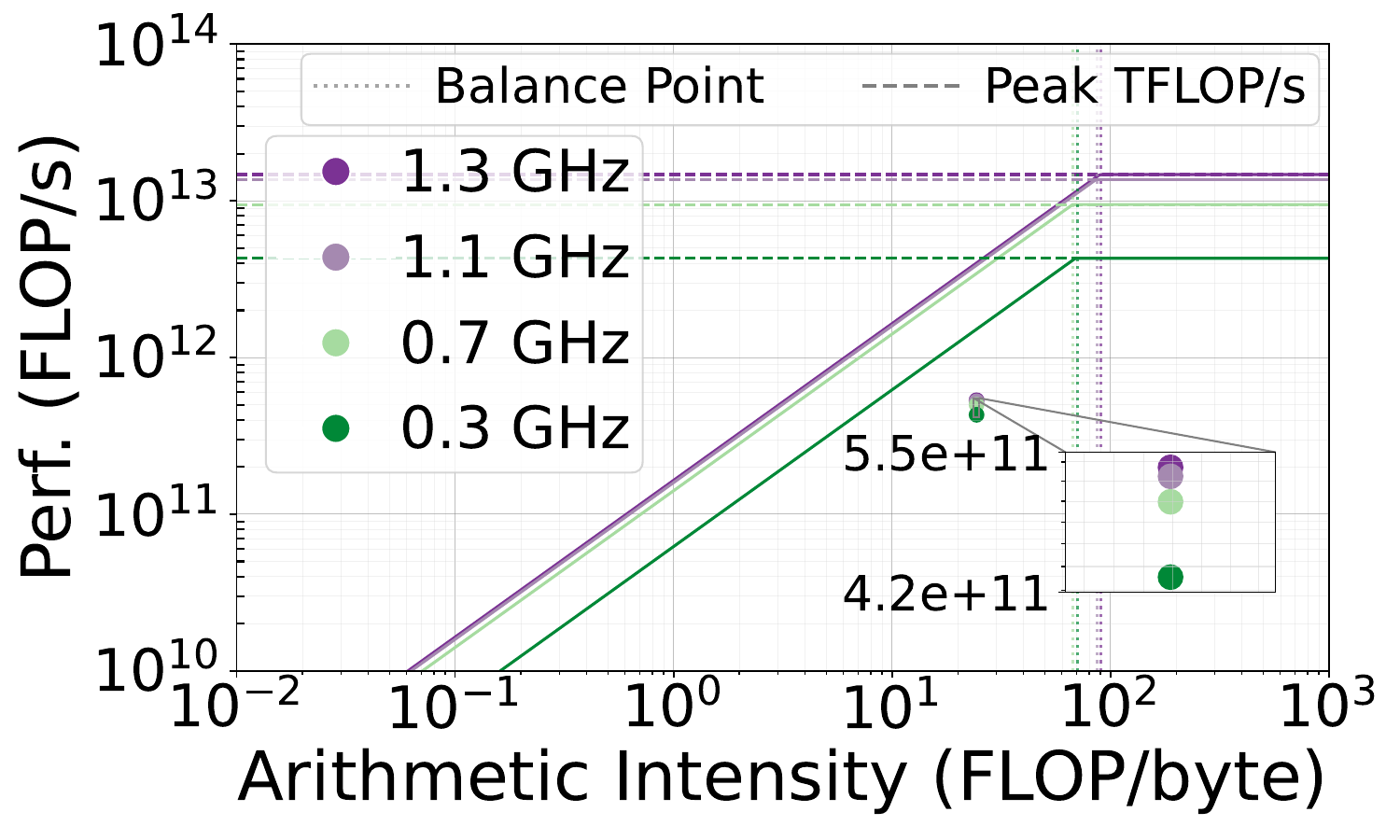}
    \label{fig:shift_time_rnetgpu}
  }
\subfloat[MobileNet]{
    \includegraphics[width=0.48\columnwidth]{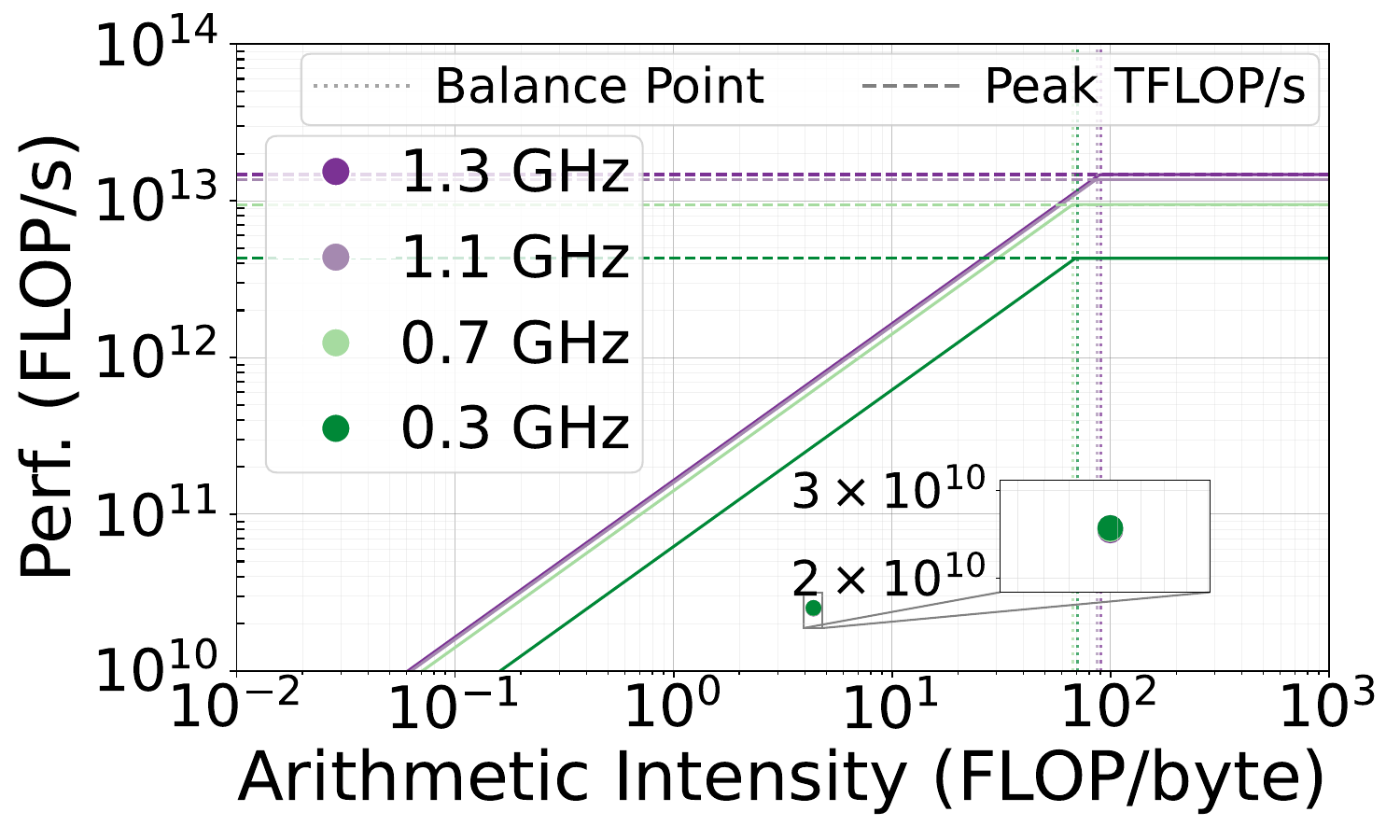}
    \label{fig:shift_time_mnetgpu}
  }

\caption{DNN Inference with shifted time roofline (GPU)} 

\label{fig:shift_time1}
\vspace{-0.15in}
\end{figure}

\begin{figure}[t]
\centering

\subfloat[ResNet]{
    \includegraphics[width=0.48\columnwidth]{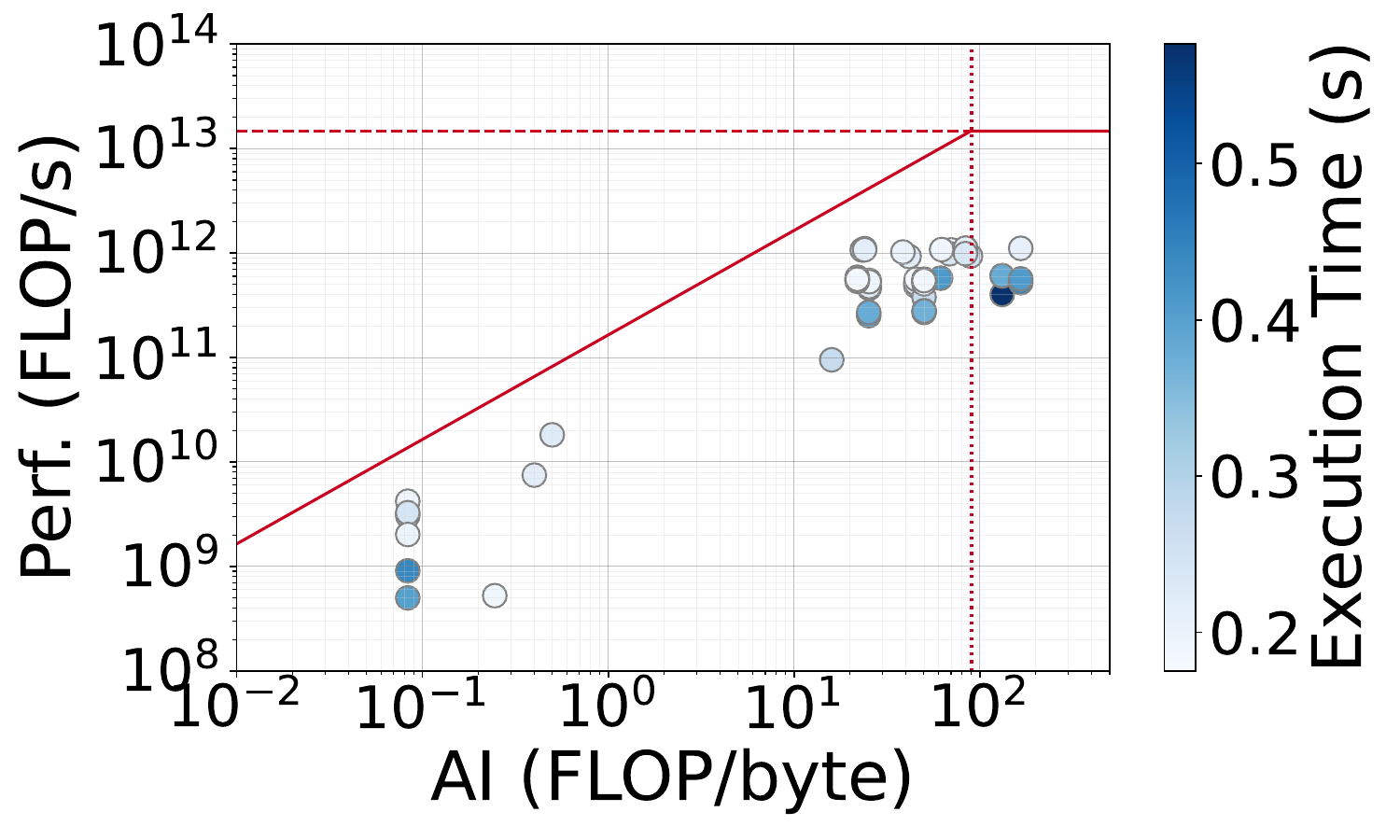}
    \label{fig:layerwise_resnet}
  }
\subfloat[MobileNet]{
    \includegraphics[width=0.48\columnwidth]{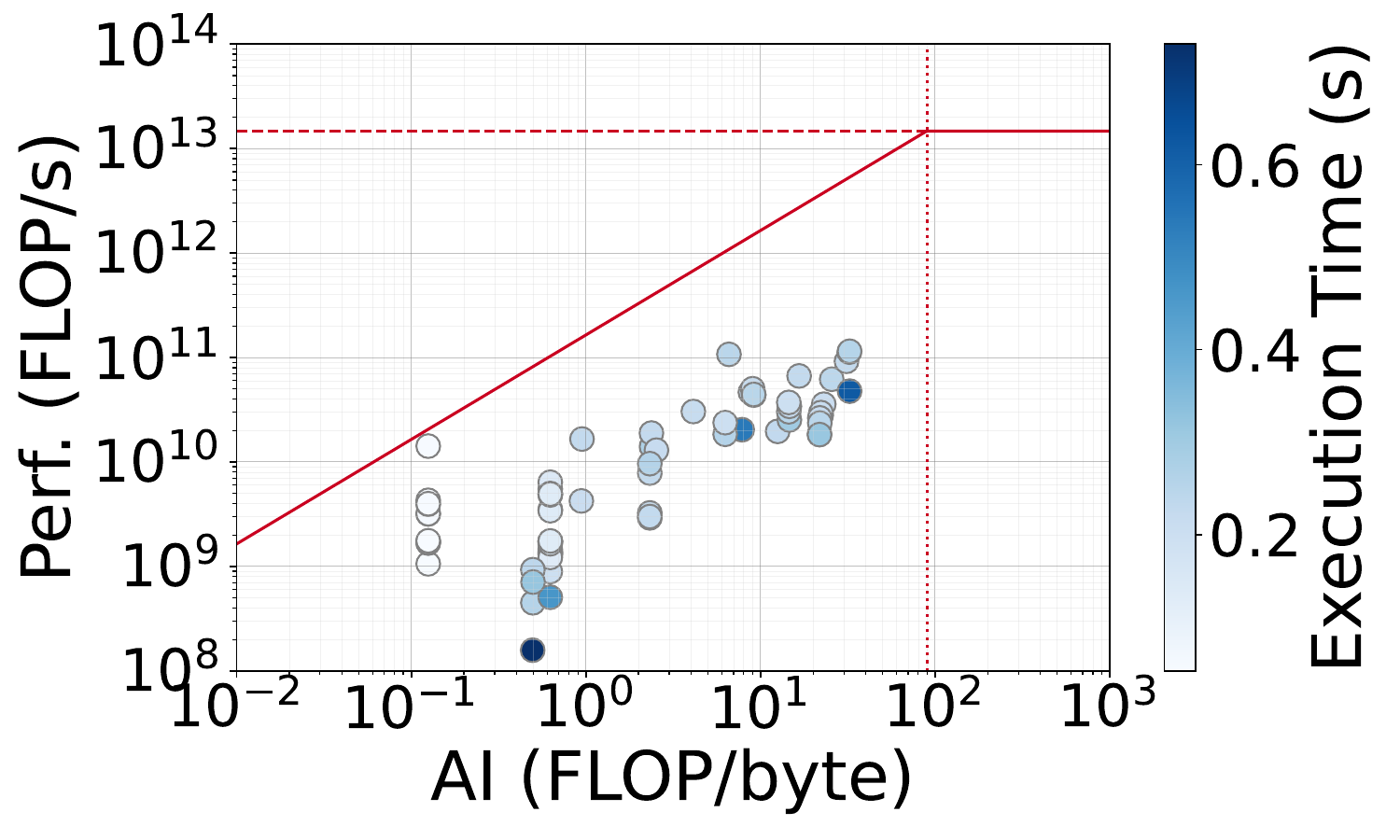}
    \label{fig:layerwise_mobilenet}
  }

\caption{Layerwise DNN Inference Performance} 
\label{fig:layerwise}

\end{figure}

\claim{Lowering memory frequency significantly degrades performance for all $5$ DNNs without any power savings}
With all other dimensions the same as MAXN, we lower the memory frequency from $3.2$GHz to $2.1$ and $0.7$GHz.  With respect to MAXN ($3.2$GHz), the time degradations for \resnet are $7.26\%$ and $41.5\%$ respectively. All $5$ DNNs see a time degradation with no power savings. This confirms expectations since all are in the memory-bound region. 

\claim{Lowering GPU frequency offers some energy reduction due to power savings, but the time degradation varies across models}
Fig.~\ref{fig:shift_time1} shows the effects of lowering GPU frequency for \resnet (high AI) and \mobilenet (low AI). Lowering GPU frequency should not degrade the performance and result in power savings since both are memory-bound. 
For \mobilenet (Fig.~\ref{fig:shift_time_mnetgpu}), when we lower the GPU frequency from $1.3$GHz to $1.1, 0.7$ and $0.3$GHz, time degrades by less than $1\%$ for all $3$ frequencies. The corresponding energy savings due to power drop are $6\%, 11.5\%$ and $15.4\%$ (not shown). This is as expected, and shows the possibility of tangible energy savings without performance degradation. However, for \resnet (Fig.~\ref{fig:shift_time_rnetgpu}), the increase in time for the $3$ GPU frequencies are $1.8\%, 7\%$ and $24\%$, with more limited energy benefits of $4\%, 10\%$ and $5\%$, respectively. 

Given the performance drop for \resnet on reducing GPU frequency even though it is memory-bound, the AI at the granularity of an entire DNN may be too coarse. We examine if layer-wise AI give us a more accurate picture.

We profile layer-wise runtimes for \resnet and \mobilenet, and use our analytical model to estimate layer-wise AI. We use these to plot layer-wise performance in Fig.~\ref{fig:layerwise}, with shade indicating the contribution of a layer to the runtime (darker is higher). 

\claim{Layer-wise AI gives a better idea of performance when the roofline is shifted}
The layer-wise AI for \mobilenet are still to the left of the time balance point, aligning with our results above of negligible performance loss when GPU frequency is lowered. For \resnet, however, several layers are compute-bound, and some of these layers significantly contribute to the runtime. This explains why its performance degrades. 

\claim{Estimating time degradation based on layer-wise AI and shifted roofline performance is non-trivial}
Based on the roofline and layer-wise AI, we reason that layers that are compute-bound will be affected when the GPU frequency is lowered. Based on Eqn.~\ref{eq:time_split}, the runtime for any layer is directly proportional to $\text{FLOPS/s}_\text{(peak)}$. So, the percentage time degradation can be calculated as the product of the $\text{FLOPS/s}_\text{(peak)}$ degradation and the percentage contribution of all compute-bound layers to the runtime. Based on this, our predicted degradations for the $3$ GPU frequencies are $1.2, 9.8$ and $17.5\%$, while the observed degradations are $1.8, 7.0$ and $24.0\%$. This shows that while the roofline can be used to analyze overall trends, it cannot directly be used as a fine-grained predictive tool. 

We also examine whether shifting the energy roofline helps in increasing the energy efficiency. We noticed mixed benefits among workloads. For instance, lowering GPU frequency to $0.7$GHz gives energy benefits of $10.8\%$ for \lstm. However, for \resnet and \mobilenet, energy efficiency did not improve. This warrants further investigation.

\section{Discussion and Conclusion}
\label{sec:conclusions}

In this article, we have presented a principled study of time and energy rooflines for DNN inferencing and training workloads for the latest, most performant edge accelerators, Nvidia Jetson AGX Orin. We analyze the effect of power modes on the time and energy rooflines, balance points and their impact on DNN inference workload performance. We also briefly discuss training workloads. Our analyses are presented as a set of insights both confirm conventional wisdom using a rigorous analytical model and empirical study, and highlight novel observations. These are useful to both deep learning researcher and practitioners.

As a practical application, we show that shifting the time roofline using power modes can optimize latency and energy. But predicting power and performance degradations based on the roofline is challenging. Although the roofline microbenchmarks show that CPU frequency and cores do not affect the peak FLOP/s and memory bandwidth, prior works~\cite{prashanthi2023sigmetrics,PowerTrain} have shown that DNN training power and performance are affected by CPU frequency due to factors like kernel launch. 

In future, we plan to incorporate CPU into the roofline model as part of future work. We also do not consider cache effects and backend optimizations such as operator fusion while modeling FLOP and memory accesses, and we plan to look into this using more sophisticated cache and GPU aware roofline techniques in future work. 

While our current focus has been on system-level factors such as power modes that can shift the roofline, there are complementary opportunities on the DNN model side. Specifically, model architecture optimizations can help increase arithmetic intensity, thereby shifting workloads from the memory-bound to the compute-bound region, e.g., by replacing memory-heavy operations with compute-intensive alternatives, fusing layers to reduce intermediate memory traffic, and reordering computations to improve data reuse~
\cite{cnninf_pmlr, abft_sc}. These can make models inherently more efficient within the roofline framework, and can complement system-level tuning.

While we focus on CNN models in this article, the recent popularity of Large Language Models (LLMs) makes extending roofline analysis to LLMs a natural progression. LLMs exhibit different computational and memory access patterns, especially due to large embedding matrices, dynamic sequence lengths, and prefill and decode phases with very different intensities~\cite{llminf_roofline}. Moreover, backend optimizations like operator fusion and quantization are more prominent in LLM inference. Since these are very large models, they often run in a distributed setting with various tuning knobs such as pipeline, tensor and expert parallelism. Extending the roofline model to distributed settings and considering communication can help choose an optimal parallelism configuration. 
    
\section*{Acknowledgments}
The authors thank students in the DREAM:Lab, IISc and Mayank Arya for their assistance. The first author was supported by the Prime Minister's Research Fellowship.

\bibliographystyle{IEEEtran}
\bibliography{arxiv}

\end{document}